\documentclass[fleqn,usenatbib]{mnras}
\usepackage{newtxtext,newtxmath}
\usepackage{longtable}
\usepackage[T1]{fontenc}
\DeclareRobustCommand{\VAN}[3]{#2}
\let\VANthebibliography\thebibliography
\def\thebibliography{\DeclareRobustCommand{\VAN}[3]{##3}\VANthebibliography}
\usepackage{subcaption}
\usepackage{graphicx}	
\usepackage{tikz}
\usepackage{blindtext}
\usepackage{placeins}
\usepackage{xcolor}

\usepackage{hyperref}

\title[106 glitches in 70 pulsars]{The Jodrell Bank Glitch Catalogue: 106 new rotational glitches in 70 pulsars}

\author[A. Basu et al.]{A. Basu,$^{1}$\thanks{E-mail: avishek.basu@manchester.ac.uk}
B. Shaw,$^{1}$\thanks{E-mail: benjamin.shaw@manchester.ac.uk}
D. Antonopoulou,$^{1}$
M. J. Keith,$^{1}$
A. G. Lyne,$^{1}$
M. B. Mickaliger,$^{1}$
\newauthor B. W. Stappers,$^{1}$
P. Weltevrede$^{1}$ and C. A. Jordan$^{1}$
\\
$^{1}$Jodrell Bank Centre for Astrophysics, School of Physics and Astronomy, University of Manchester, Manchester, UK, M13 9PL\\
}

\date{Accepted XXX. Received YYY; in original form ZZZ}

\pubyear{2020}

\begin{document}
\label{firstpage}
\pagerange{\pageref{firstpage}--\pageref{lastpage}}
\maketitle

\begin{abstract}
Pulsar glitches are rapid spin-up events that occur in the rotation of neutron stars, providing a valuable probe into the physics of the interiors of these objects. Long-term monitoring of a large number of pulsars facilitates the detection of glitches and the robust measurements of their parameters. The Jodrell Bank pulsar timing programme regularly monitors more than 800 radio pulsars and has accrued, in some cases, over 50 years of timing history on individual objects. In this paper we present 106 new glitches in 70 radio pulsars as observed up to the end of 2018. For 70\% of these pulsars, the event we report is its only known glitch. For each new glitch we provide measurements of its epoch, amplitude and any detected changes to the spin-down rate of the star. Combining these new glitches with those listed in the Jodrell Bank glitch catalogue we analyse a total sample of 543 glitches in 178 pulsars. We model the distribution of glitch amplitudes and spin-down rate changes using a mixture of two Gaussian components. We corroborate the known dependence of glitch rate and activity on pulsar spin-down rates and characteristic ages, and show that younger pulsars tend to exhibit larger glitches. Pulsars whose spin-down rates between $10^{-14}$ Hz s$^{-1}$ and $10^{-10.5}$ Hz s$^{-1}$ show a mean reversal of 1.8\% of their spin-down as a consequence of glitches. Our results are qualitatively consistent with the superfluid vortex unpinning models of pulsar glitches. 
\end{abstract}

\begin{keywords}
stars: neutron -- pulsars: general -- methods: statistical -- methods: data analysis
\end{keywords}

\section{Introduction}\label{introduction}
The technique of pulsar timing is used for the precise measurement of the rotational and astrometric parameters of pulsars. Frequent monitoring of a sample of pulsars allows this technique to be used to study their long-term spin evolution. The most noticeable feature is the steady spin-down, resulting from the loss of rotational energy in the form of particle winds and electromagnetic radiation. This steady spin-down is occasionally interrupted by spin-up events, observed as a discontinuous change in the rotation frequency called glitches \citep{Radman+1969,reichleidowns69}. Following a glitch, there may be a period of recovery of the spin frequency towards pre-glitch values, often in an exponential manner (e.g.,  \citealt{Yu+2012}, \citealt{aeka18}, \citealt{Basu+2019}). However, the recovery may be interrupted by the occurrence of subsequent glitches, resulting in a cumulative effect on the long-term spin-down of the pulsar \citep{Lyne+1996, antonopoulou18}. Glitches and their recoveries are understood to arise due to the interaction of an internal superfluid component of the neutron star (NS) with the non-superfluid component comprising the ion lattice in the crust and charged part of the core \citep{Baym+1969,AndersonItoh1975}. The rotating superfluid forms an array of vortices, the areal density (number of vortices per unit area) of which determines its angular velocity. Pinning of vortices to the crustal lattice sites can prevent spin-down of the superfluid along with  the rest of the star (the normal component), which gives rise to a velocity lag between the two and a build-up of angular momentum excess in the superfluid. When vortices unpin from the lattice at large, they transfer some, or all, of the angular momentum supply to the outer crust \citep{AndersonItoh1975} spinning it up rapidly; this is  observed as a glitch.  
The underlying mechanism responsible for triggering large scale vortex unpinning required to produce glitches is not yet well understood \citep[see, e.g., the review by][]{haskellmelatos15}.

The observed change in the rotational frequency ($\Delta \nu$), spans $\sim 6$ orders of magnitude. The time interval between two consecutive glitches in combination with $\Delta \nu$  is a useful quantity to estimate the fraction of moment of inertia of the superfluid component that contributes to a glitch \citep{Link+1999,  anderson+2012}. 
The canonical models for pulsar glitches associate the angular momentum reservoir with the superfluid in the neutron star's crust \citep{Alpar+1977}. However, the largest glitches suggest the presence of a reservoir extending beyond the crust. When the effects of entrainment\footnote{The nuclear lattice in the crust inhibits the free flow of superfluid neutrons, reducing the moment of inertia of the ``free" crustal superfluid.  This effect is described by an \emph{entrainment} parameter \citep[see e.g.][]{Chamel+2017}.} \citep{Chamel+2012} are taken into account, a small fraction of glitches require contribution from the core superfluid \citep{Nchamel2013, anderson+2012, basuFMI}.

High cadence monitoring of the pulse times of arrival (ToAs) helps to constrain and  parameterise the glitch rise and any subsequent recovery, which  provide inputs to theoretical models of the coupling between the superfluid and normal components. Observed post-glitch relaxation is due to superfluid components within the neutron star that respond to the spin-up on local timescales, defined by their coupling strength \citep{AlparvortexcreepI1984, Jones+2002, Anderson+Sidery+Comer+2006,Haskell+2012, Newton+2015}. The strong coupling regime which defines the spin-up rise timescale \citep{Epstein+Baym+1992, Jones+1992} can be probed when a glitch occurs during the observations and is resolved in time. 
The first observation of the spin up in the Vela pulsar's 2000 glitch put an upper limit of 40\,s on the rise timescale \citep{Dodson+2002}. The third such resolved spin-up in the Vela pulsar was in 2016  \citep{pal16} and was constrained to occur in less than 12.5 s by \citet{GA+2019}. Larger glitches in the Crab pulsar have been shown to exhibit slow rises occurring on timescales up to $\sim$2 days  (e.g., \citealt{Shaw+2018, Basu+2019, Ge+2020, Bshaw+2021}). Such observations provide crucial information to the theoretical models \citep{Graber+2018, HKAD+2018, Montolli+2020} and, together with precise measurements of the glitch parameters and recovery, advance our understanding of the micro-physics of the NS interior. 

In addition to glitches, many pulsars exhibit another form of timing irregularity known as timing noise, characterised by a quasi-random wandering of the rotational parameters.
Although the mechanism driving timing noise is not well understood, a number of explanations have been suggested. These include instabilities in the pulsar magnetosphere \citep{Kramer+2006, Lyne+Hobbs+2010} and the existence of a turbulent superfluid component inside the NS \citep{Melatos+Link+2014}. Typically, timing noise is modelled as a red-noise process (see \S \ref{method}). It is important to account for timing noise in the modelling of pulsar rotation, in order to obtain a robust estimate of the glitch parameters.

Two large-scale studies of glitches have been previously undertaken. \cite{espinoza2011} presented a detailed study of 315 glitches in 102 pulsars, and \cite{Yu+2012} analysed 107 glitches and their recoveries in 36 pulsars. In this paper we present 106 new glitches in 70 pulsars, observed over 7 years of extended monitoring since \cite{espinoza2011} using the telescopes at the Jodrell Bank Observatory (JBO). In \S \ref{observations} we briefly discuss the ongoing observational program, which to date has accrued $\sim$17000 years of pulsar rotation history. In \S \ref{method} we explain the method adopted for measuring the glitch parameters, whilst the updated glitch database, general properties and a statistical analysis of glitches are discussed in \S \ref{table_G}. We finally conclude the paper by summarising our findings from this work in \S \ref{conclude}. All measurements  are available through the Jodrell Bank pulsar glitch catalogue\footnote{\href{http://www.jb.man.ac.uk/pulsar/glitches/gTable.html}{http://www.jb.man.ac.uk/pulsar/glitches/gTable.html}}.


\section{Observations}\label{observations}
\begin{figure}
    \centering
    \includegraphics[scale=0.52]{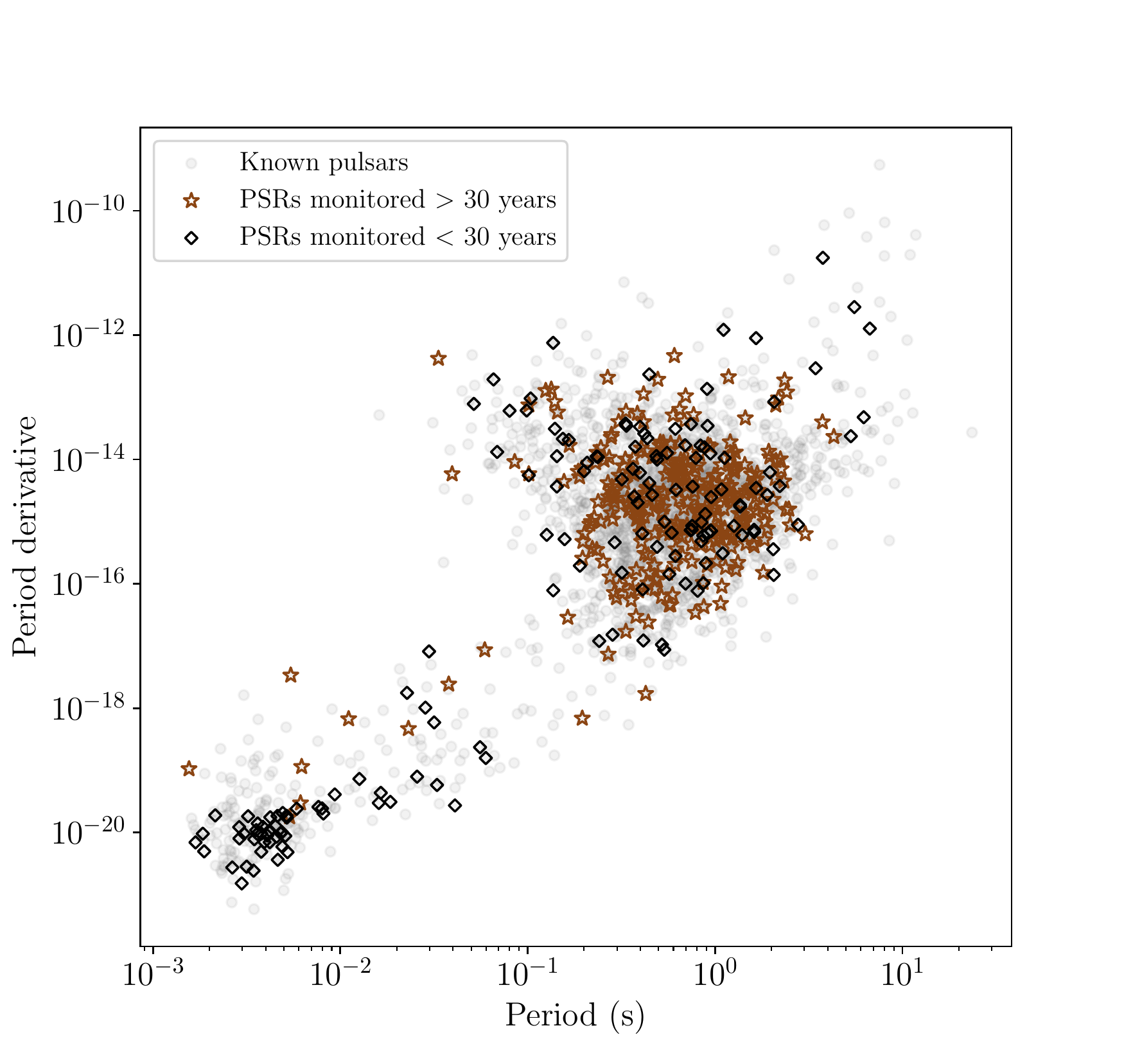}
    \caption{ The $P-\dot{P}$ diagram for all known pulsars, highlighting those monitored at the JBO. The dots represents all the known pulsars listed in the ATNF catalogue (accessed on 2021 April 1), the open stars indicate the pulsars which have been monitored for more than 30 years at JBO and the open diamonds indicate those which have been monitored for less than 30 years.}
    \label{fig:Jodrell_psr}
\end{figure}

The pulsar timing programme at JBO routinely monitors the rotation of $\sim$800 pulsars, primarily using the 76-m Lovell telescope. A number of supplementary observations are carried out using the 38$\times$25-m Mark-II telescope, also at JBO. In addition, we routinely monitor a small number of sources with the 42-ft JBO telescope. Figure \ref{fig:Jodrell_psr} presents the distribution of the $\sim$ 800 pulsars observed at JBO in a period $P$-period derivative $\dot{P}$ diagram. For a given pulsar, the typical interval between observations ranges from daily to monthly, but for most pulsars it is $\sim$10 to 14 days. Typical integration times are in the range $\sim$6 to 30 minutes. 

Lovell and Mark II observations are carried out according to the procedures outlined in \cite{hlk+04}, \cite{sl96}  and \cite{mhb+13}. To summarise, data prior to 2009 are recorded using an analogue filterbank (AFB) and a digital filterbank (DFB) was used thereafter. For the AFB observations, a bandwidth of 32 MHz was used, centered on 1400 MHz with a small number of observations carried out at  500, 600 or 925 MHz. For DFB observations, a bandwidth of 400 MHz, centred on 1520 MHz was used.  Data are recorded using 400 or 512 bins per pulse period (AFB) or 1024 bins (DFB), written as 1-minute (AFB) or 10 s (DFB) sub-integrations over 256 (AFB) or 768 (DFB) frequency channels. For millisecond pulsars, data are folded onto 512 bins (AFB), or 64, 128, 256, 512 or 1024 bins (DFB) \citep[see][]{IPTA-DR2}. The effects of interstellar dispersion are removed using incoherent dedispersion in hardware and the data are folded online according to the topocentric period of the pulsar. Each sub-integration is formed by the summation of all of the individual pulses that were recorded over the sub-integration duration. 

The 42-ft telescope observations are carried out using a bandwidth of 5-10 MHz at frequencies near 610 MHz using an AFB or COBRA2 backend. Data are recorded using 512 (AFB) or 1024 (COBRA2) bins per pulse period written as 10-minute (AFB) or  1-minute (DFB) long sub-integrations. AFB data were dedispersed incoherently in hardware whereas coherent dedispersion \citep{hr75} was used for the COBRA2 data in software. Folding and sub-integration formation is carried out as described above.  

The removal of radio frequency interference (RFI) is a multi-stage process. A median filtering algorithm is used initially to remove the RFI from the DFB data. The remaining RFI, observed predominantly as bad frequency channels, but sometimes entire sub-integrations, is removed manually. For the AFB data, only the second stage is applied. 

The remaining sub-integrations, frequency channels and polarisations are averaged to form a single integrated profile for each observation using {\sc psrchive} \citep{sdo12}. In order to determine the pulse time-of-arrival (ToA), the integrated profile is then cross-correlated in the time-domain\footnote{ Compared to frequency domain techniques, this method has been shown to improve ToA precision for observations with low signal-to-noise ratios}, with a noise-free template profile, formed by fitting von-Mises functions to a high signal-to-noise ratio observation of the pulsar, which represents and idealised realisation of the expected pulse profile.  The measured ToAs are then corrected to the approximately inertial Solar system barycentre, using the DE436 planetary ephemeris\footnote{\href{https://naif.jpl.nasa.gov/pub/naif/JUNO/kernels/spk/de436s.bsp.lbl}{https://naif.jpl.nasa.gov/pub/naif/JUNO/kernels/spk/de436s.bsp.lbl}}, forming Barycentric arrival times, which are then used for the timing analysis which is discussed below. 

\section{Methodology}\label{method}

\begin{figure*}
\centering
\begin{subfigure}{.5\textwidth}
    \centering
    \includegraphics[width=1\textwidth]{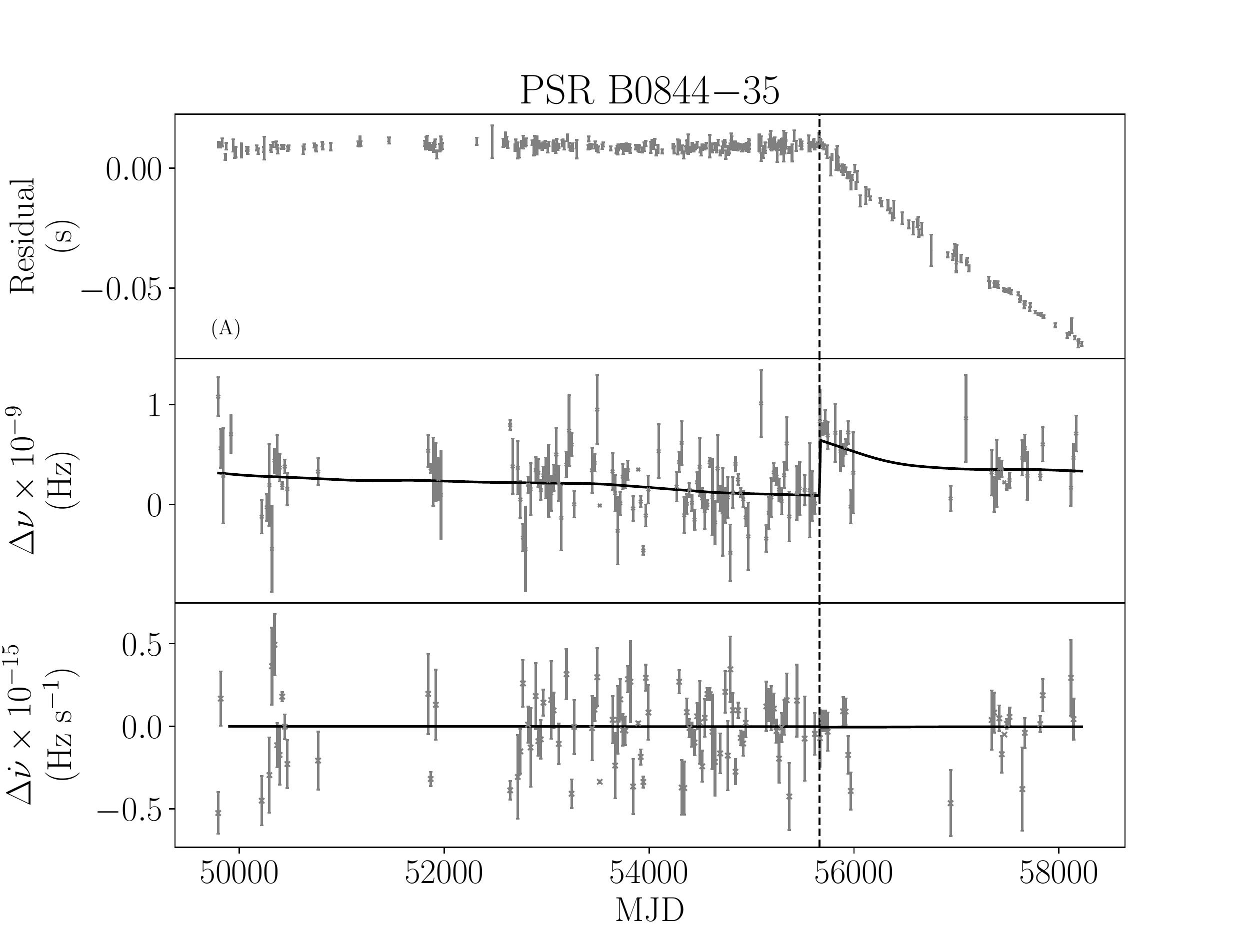}
\end{subfigure}%
\begin{subfigure}{.5\textwidth}
    \centering
    \includegraphics[width=1\textwidth]{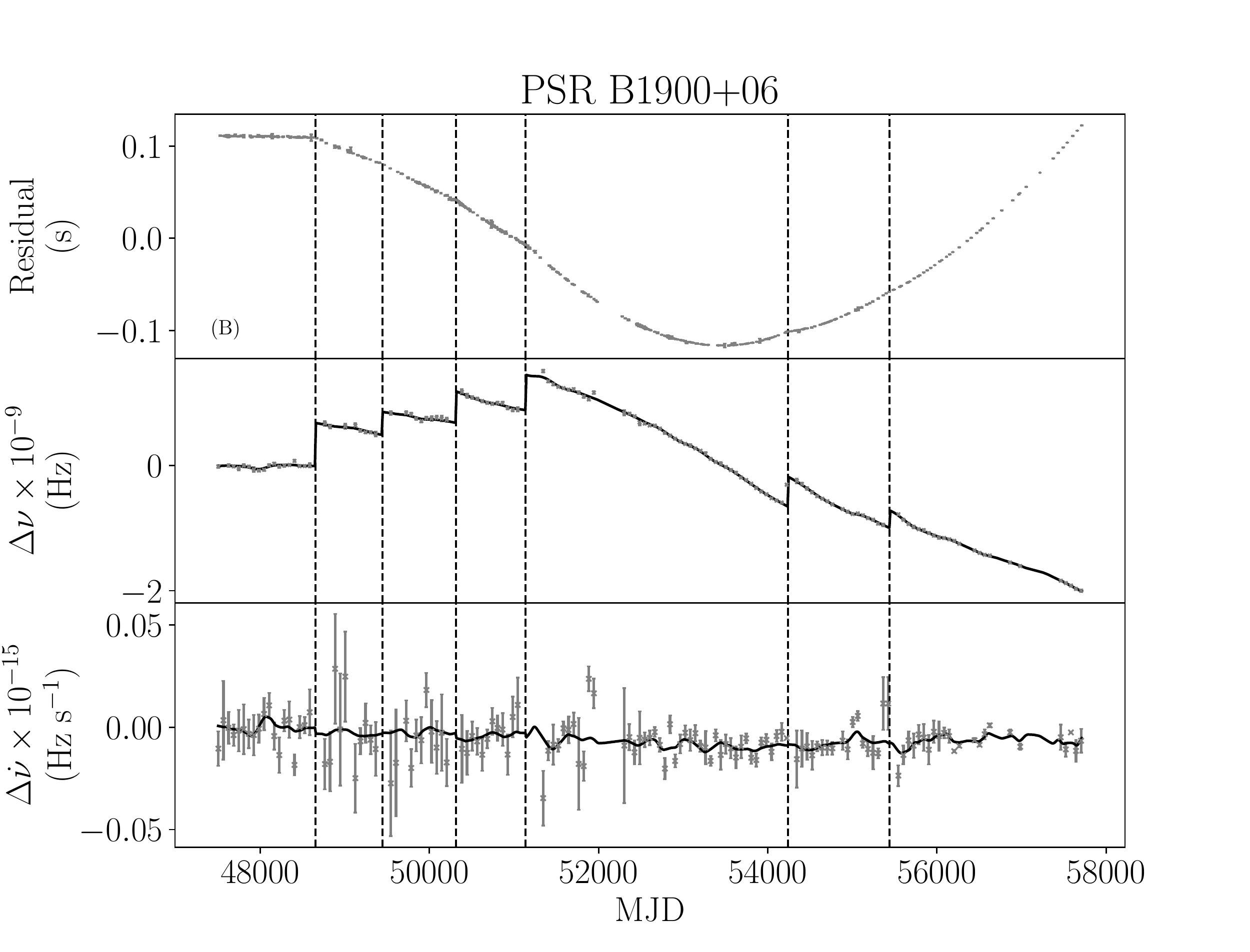}
\end{subfigure}
\begin{subfigure}{.5\textwidth}
    \centering
    \includegraphics[width=1\textwidth]{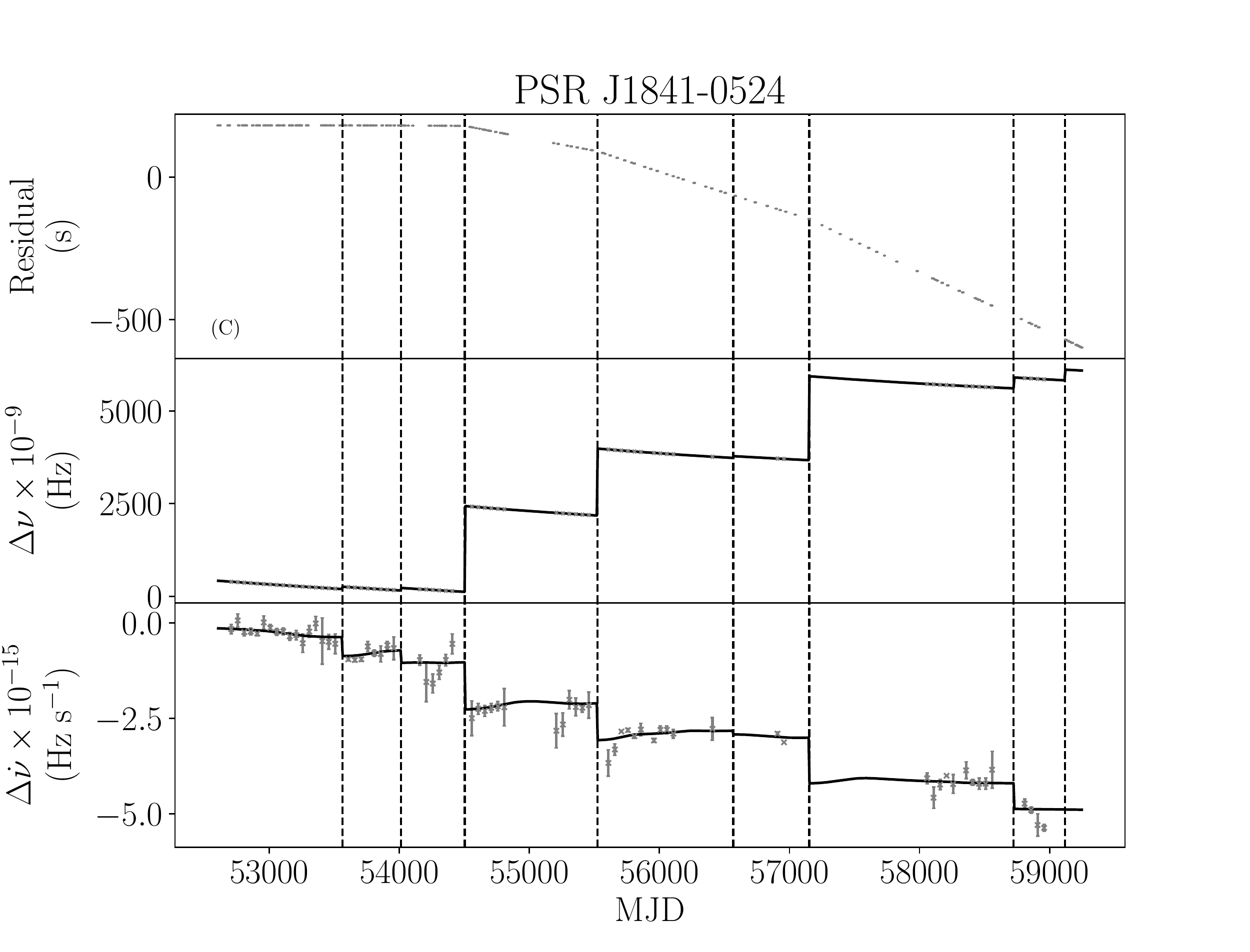}
\end{subfigure}%
\begin{subfigure}{.5\textwidth}
    \centering
    \includegraphics[width=1\textwidth]{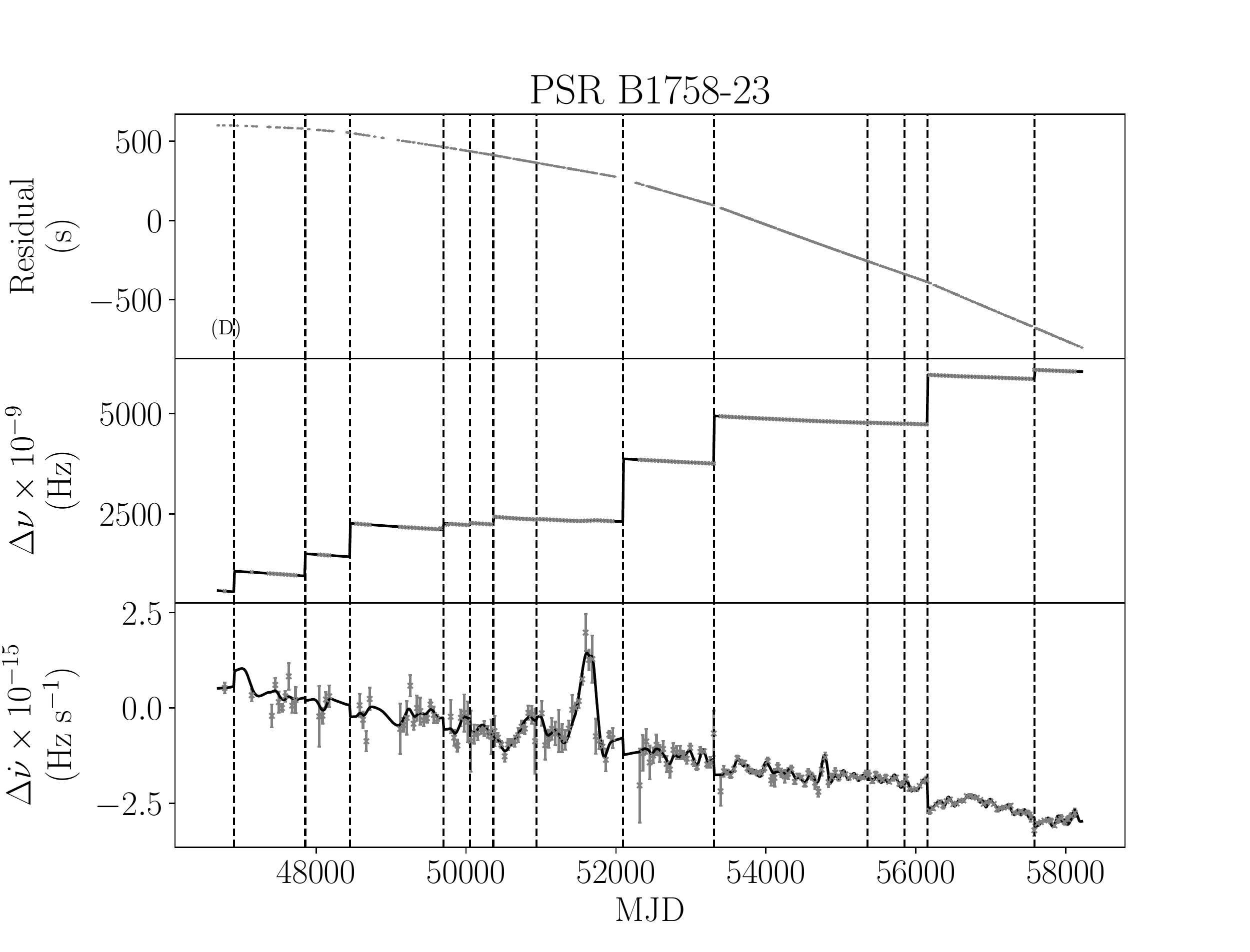}
\end{subfigure}
\caption[short]{The timing residuals, $\Delta \nu$ and $\Delta \dot \nu$ as a function of time for four different pulsars, which demonstrate their differences in rates of glitch occurrence, and glitch amplitude (see text). The points represent the rotational evolution obtained using a striding boxcar method (see \citealp{Shaw+2018}) and the solid lines indicate the solutions from a Gaussian process regression analysis (see text). The vertical dashed lines denote glitch epochs.}
 \label{fig:residual_plots}
\end{figure*}

\subsection{Pulsar timing}\label{timing}
In pulsar timing the temporal evolution of the rotation phase ($\phi$) is modelled using the Taylor series expansion 
\begin{equation}\label{simple_model}
\phi(t) =\phi_0 + \nu_0(t-t_0) + \frac{1}{2}\dot{\nu}_0(t-t_0)^2 + \frac{1}{6}\Ddot{\nu}_0(t-t_0)^3,
\end{equation}
where $\nu_0$, $\dot\nu_0$ and $\ddot\nu_0$ are the rotation frequency, and its first and second time derivatives at some fiducial epoch $t_0$. Here $t$ represents the measured barycentric pulse arrival times.
The initial phase $\phi_0$ is arbitrary and can be chosen to be zero at time $t=t_0$, without any loss of generality. Since the value of $\ddot\nu_0$ is very small, often the terms up to first order in rotation frequency are sufficient to model the rotational phase over a long period of time. The integer part of $\phi$ for a given ToA represents the number of turns of the pulsar since $t=t_0$. The difference between $\phi$ and the closest integer to $\phi$ is defined as a phase residual, which is minimised by fitting for the rotational parameters in Equation \ref{simple_model} as well as any astrometric, interstellar medium (ISM), and (where applicable) binary parameters. The residuals therefore represent the differences between measured ToAs and those predicted by the model. 

For each pulsar in our sample, we produce the timing model that approximately describes the rotational phase of each pulsar over the full data span. This model is then used to calculate the residuals. Phase ambiguities are resolved by means of tracking the integer number of rotations between each ToA using the pulse numbering feature of \textsc{tempo2} \citep{tempo2}. These pulse numbers have been computed by careful inspection of each individual data set to smoothly track the rotational phase of the pulsar over the entire data set (coherent timing). 

In the presence of a  glitch, the residuals following the event typically display a systematic pattern like those shown in Figure \ref{fig:residual_plots}. In the left hand top panel we show the residuals for PSR B0844$-$35 which is an example of an infrequent glitcher whereas PSR B1900$+$06 (right hand top panel) exhibits regular small glitches. The lower left hand panels show PSR J1841$-$0524 and the lower right hand panel shows PSR B1758$-$23. Both of these are frequent glitchers but PSR J1841$-$0524 exhibits glitches of a characteristic size whereas those of PSR B1758$-$23 span a wide range of sizes. These deviations can be modelled phenomenologically by assuming the glitch induced phase evolution ($\phi_g$), which sets in after the glitch epoch $t_g$, can be expressed as
\begin{eqnarray} \label{glitch_evolution}
    \phi_g(t) = \Delta \phi + \Delta \nu_{\text{p}} (t-t_g) + \frac{1}{2}\Delta \dot{\nu}_{\text{p}} (t-t_g)^2 + \\ \sum_i \Delta \nu_{\text{d}_i}\tau_i(1-e^{{-(t-t_g)}/{\tau_i}}). \nonumber
\end{eqnarray}
Post-glitch, the total phase evolution is $\phi(t) + \phi_g(t)$. In Equation \ref{glitch_evolution}, $\Delta \nu_{\text{p}}$ and $\Delta \dot{\nu}_{\text{p}}$ are respectively the persistent step changes in spin frequency and spin-frequency derivative.
In a  number of cases the post-glitch evolution displays either a single or multiple exponential recovery where the component $\Delta \nu_{\text{d}}$ recovers exponentially with a timescale denoted by $\tau$. The quantity $\Delta \phi$ is included to account for any discontinuity in phase because of uncertainty in the calculation of the glitch epoch (see appendix \ref{Apn1}). Hence, from the glitch model in Equation \ref{glitch_evolution}, the change in rotation frequency at the time of glitch is $\Delta \nu = \Delta \nu_{\text{p}} + \sum_i \Delta \nu_{\text{d}_{i}}$ and the change in its first derivative is $\Delta \dot{\nu} = \Delta \dot{\nu}_{\text{p}} + \sum_i \Delta \nu_{\text{d}_{i}}/\tau_i$.

Glitches are typically identified by manual inspection of the timing residuals for each pulsar. The first post-glitch ToA will arrive early with respect to the pre-glitch timing model, allowing an initial estimate of the glitch epoch. We fit a pulsar timing model to the ToAs that includes the spin frequency and two derivatives ($\nu_0$,$\dot\nu_0$, $\ddot\nu_0$), astrometric parameters (position, proper motion and parallax), a glitch model and a Gaussian process model of the red spin noise using a power-law Fourier basis. We also fit for additional, or poorly estimated, white noise with the commonly used EFAC and EQUAD parameters \citep{lah+13}.
The Gaussian process red noise model is parameterised by the log-amplitude, $\log_{10}(A_\mathrm{red})$, and slope, $\alpha$, of a power-law in the power-spectral density, as described in \citet{lah+13}. The power-spectral density of the red noise at a fluctuation frequency $f$ is given by
\begin{equation}
    P(f) = \frac{A_\mathrm{red}^2}{12\pi^2} \, f^{-\alpha}.
\end{equation}
For each pulsar all timing model parameters, including those for all  glitches, are fitted simultaneously. The red noise 
and white noise parameters are sampled using a Markov-chain Monte-Carlo (MCMC) approach, and the other parameters are marginalised over using the linearised approximation to the timing model generated by \textsc{tempo2}; the likelihood is computed using \textsc{enterprise} \citep{evt+19}. Sampling is performed using \textsc{emcee} \citep{2013ForemanMackey}. Glitch parameters are then extracted from the maximum likelihood solution using \textsc{tempo2}.

In this paper we only present the results from our timing analysis pertaining to glitches. A full study of the red spin noise, spin evolution and astrometric parameters will appear in future works.

\section{Results}\label{table_G}
\begin{table}
  \begin{center}
\caption{Table showing the measured glitch parameters from all of the new glitches in our sample. The first and the second columns list the J-name and B-name of the pulsar. The third column indicates the glitch epoch and its uncertainty (in parentheses). The fourth and the fifth columns indicate the relative changes in the rotation frequency and its first order derivative. With the exception of the glitch epoch (see Appendix \ref{Apn1}), all uncertainties are quoted at the 2$-\sigma$ level.}
    \label{glitch_table}
    \begin{scriptsize}
    \begin{tabular}{lllll}
      \hline
      \hline
      \multicolumn{1}{c}{PSR}
      & \multicolumn{1}{c}{PSR}
      & \multicolumn{1}{c}{MJD} 
      & \multicolumn{1}{c}{$\Delta{\nu}/\nu$} 
      & \multicolumn{1}{c}{$\Delta{\dot{\nu}}/\dot{\nu}$} \\
      \multicolumn{1}{c}{J$-$name}
      & \multicolumn{1}{c}{B$-$name}
      & \multicolumn{1}{c}{days}
      & \multicolumn{1}{c}{$10^{-9}$}
      & \multicolumn{1}{c}{$10^{-3}$}\\
      \hline 
      \hline 
J0157$+$6212 & B0154+61 & 58283(3)     & 2.6(3)      & $--$  \\
J0205+6449 & $--$        & 54907(18)    & 1761(7)     & 12(2)     \\
J0205+6449 & $--$       & 55719(10)    & 126(5)      & 3(1)      \\
J0205+6449 & $--$        & 55836(5)     & 2933(5)     & 13(1)     \\
J0205+6449 & $--$        & 56854(28)    & 84(14)    & 3(2)      \\[4.5pt]
J0205+6449 & $--$        & 57345(12)    & 527(7)      & 5(1)      \\
J0205+6449 & $--$        & 57696.75(8)  & 43(2)       & $--$      \\
J0205+6449 & $--$        & 58320(97)    & 3140(50)    & 20(3)     \\
J0215+6218 & $--$        & 55965(27)    & 0.06(4)     & $--$   \\
J0415+6954 & B0410+69 & 57120(60)    & 0.011(6)    & $--$   \\[4.5pt]
J0417+35   & $--$        & 57053.63(6)  & 24.54(3)    & 3.4(1)    \\
J0525+1115 & B0523+11 & 56010.4(5)   & 0.312(2)    & 0.6(2)    \\
J0534+2200 & B0531+21 & 55875.49(3)   & 39(4)      & 0.23(2)   \\
J0534+2200 & B0531+21 & 57839.8(1)    & 2.2(4)     & $--$         \\
J0534+2200 & B0531+21 & 58470.7(2)    & 2.3(6)     & $--$        \\[4.5pt]
J0601$-$0527 & B0559$-$05 & 54347(13)    & 0.05(1)     & 0.2(2)    \\
J0611+1436 & $--$        & 55818(16)    & 5575.4(8)   & $--$         \\
J0625+1015 & $--$       & 55203(8)     & 0.33(5)     & $--$         \\
J0625+1015 & $--$        & 57676(3)     & 1.17(6)     & 0.9(2)    \\
J0631+1036 & $--$        & 55119.35(6)  & 9.4(4)      & 0.2(5)    \\[4.5pt]
J0631+1036 & $--$        & 55276.7(3)   & 7.5(6)      & $--$         \\
J0631+1036 & $--$        & 55702(5)     & 3277.9(8)   & 2.7(6)    \\
J0631+1036 & $--$        & 58341.7(1)   & 27.8(8)     & $--$    \\
J0631+1036 & $--$        & 58352.13(1)  & 96.4(6)     & 1.3(3)    \\
J0742$-$2822 & B0740$-$28 & 56725.2(2)   & 2.4(1)      & $-$1.1(6)   \\[4.5pt]
J0820$-$1350 & B0818$-$13 & 49927(20)    & 0.035(2)    & 0.159(4)  \\
J0820$-$1350 & B0818$-$13 & 58063(10) & 0.082(6)       & 0.14(7)   \\
J0846$-$3533 & B0844$-$35 & 55679(5)     & 0.62(4)     & 3(1)      \\
J0922+0638 & B0919+06 & 55152(8)     & 1256.2(4)   & 0.1(9)    \\
J1635$-$2614 & $--$        & 56472(3)     & 0.71(3)     & 0.1(4)    \\[4.5pt]
J1705$-$1906 & B1702$-$19 & 55171(3)     & 0.28(3)     & $--$         \\
J1705$-$1906 & B1702$-$19 & 55955(1)     & 0.30(3)     & $--$         \\
J1705$-$1906 & B1702$-$19 & 57954(10)    & 0.18(5)     & $--$         \\
J1730$-$3350 & B1727$-$33 & 55908(30)    & 2254(1)     & 5.3(5)    \\
J1734$-$3333 & $--$        & 56340(4)     & 90(10)      & $--$   \\[4.5pt]
J1740$-$3015 & B1737$-$30 & 55213(75)    & 2663(3)     & 1.2(3)    \\
J1740$-$3015 & B1737$-$30 & 55936.4(2)   & 18.3(6)     & 0.4(2)    \\
J1740$-$3015 & B1737$-$30 & 57468.599(5) & 229.4(4)    & 1.2(4)    \\
J1740$-$3015 & B1737$-$30 & 58241(8)     & 837.9(7)    & 2.0(3)    \\
J1737$-$3102 & $--$        & 57087(2)     & 18(1)       & $--$         \\[4.5pt]
J1737$-$3137 & $--$   & 57115(3)     & 7.0(7)      & $--$         \\
J1737$-$3137 & $--$        & 58147(42)    & 4498(1)     & 1.3(1)    \\
J1740+1000 & $--$        & 56164(10)    & 2910.9(4)   & 11(1)     \\
J1746$-$2856 & $--$        & 57038(1)     & 24.4(4)     & 2.1(3)    \\
J1751$-$3323 & $--$        & 57019(1)     & 13(1)       & $-$5(2)     \\[4.5pt]
J1757$-$2421 & B1754$-$24 & 55716(21)    & 7807(1)     & 6.3(5)    \\
J1801$-$2451 & B1757$-$24 & 56979(36)    & 2413(4)     & 5.3(8)    \\
J1801$-$2304 & B1758$-$23 & 55360(3)     & 2.9(7)      & $--$         \\
J1801$-$2304 & B1758$-$23 & 55841(3)     & 1.7(5)      & $--$        \\
J1801$-$2304 & B1758$-$23 & 56149.024(8) & 513.8(4)    & 1.2(7)    \\[4.5pt]
J1801$-$2304 & B1758$-$23 & 57586.2(1)   & 96.6(9)     & $--$         \\
J1809$-$2004 & $--$        & 56227.5(2)   & 48.67(5)    & 0.79(3)   \\
J1814$-$1744 & $--$        & 55626(2)     & 16(4)       & 0.3(2)    \\
J1827$-$0958 & B1824$-$10 & 48764(18)    & 0.4(1)      & $--$        \\
J1835$-$1020 & $--$        & 57589.5(3)   & 3.26(2)     & 0.2(1)    \\[4.5pt]
J1826$-$1334 & B1823$-$13 & 56556(4)     & 2599.7(3)   & 10.1(4)   \\
J1830$-$1059 & B1828$-$11 & 55040.9(2)   & 6.3(5)      & $--$         \\
\hline \multicolumn{5}{r}{{\dots \sl continued}}\\
\end{tabular}
\end{scriptsize}
\end{center}
\end{table}

\begin{table}
  \begin{center}
    \begin{scriptsize}
    \contcaption{}
    \begin{tabular}{lllll}
      \hline
      \hline
      \multicolumn{1}{c}{PSR}
      & \multicolumn{1}{c}{PSR}
      & \multicolumn{1}{c}{MJD} 
      & \multicolumn{1}{c}{$\Delta{\nu}/\nu$} 
      & \multicolumn{1}{c}{$\Delta{\dot{\nu}}/\dot{\nu}$} \\
      \multicolumn{1}{c}{J$-$name}
      & \multicolumn{1}{c}{B$-$name}
      & \multicolumn{1}{c}{days}
      & \multicolumn{1}{c}{$10^{-9}$}
      & \multicolumn{1}{c}{$10^{-3}$}\\
      \hline \hline
J1833$-$0827 & B1830$-$08 & 54779(2)     & 0.5(1)      & 0.15(8)   \\
J1836$-$1008 & B1834$-$10 & 55744.5(3)   & 3.6(1)      & $--$         \\
J1837$-$0604 & $--$        & 55821(23)    & 1398(1)     & 6.9(7)    \\[4.5pt]
J1837$-$0604 & $--$        & 56511.0(4)   & 16.7(8)     & $--$         \\
J1837$-$0559 & $--$        & 56258(52)    & 0.2(2)      & $--$         \\
J1837$-$0559 & $--$        & 58022(4)     & 3.4(3)      & $--$         \\
J1839$-$0223 & $--$        & 56994(2)     & 3.17(4)     & 3.5(2)    \\
J1841$-$0524 & $--$        & 55525(35)    & 804.9(4)    & 0.8(1)    \\[4.5pt]
J1841$-$0524 & $--$        & 56564.6(7)   & 21.6(4)     & $--$         \\
J1844$-$0433 & B1841$-$04 & 49358(9)     & 0.22(1)     & $-$0.03(2)  \\
J1841$-$0157 & $--$        & 56923.6(4)   & 33.3(4)     & $--$         \\
J1842+0257 & $--$       & 57112(55)    & 12060.9(2)  & 1.0(3)    \\
J1844+00   & $--$        & 56784(43)    & 249.55(3)   & 4.1(7)    \\[4.5pt]
J1845$-$0743 & $--$        & 57939(23)    & 0.04(1)     & $-$0.11(9)  \\
J1847$-$0402 & B1844$-$04 & 55502(2)     & 0.52(4)     & 0.04(3)   \\
J1849$-$0317 & $--$        & 58306(4)     & 33(2)       & $--$         \\
J1850$-$0026 & $--$        & 56239.2(5)   & 14.2(3)     & 0.9(2)    \\
J1850$-$0026 & $--$        & 56425(11)    & 91.9(3)     & 0.8(2)    \\[4.5pt]
J1850$-$0026 & $--$        & 57495.3(6)   & 7.0(2)      & 0.7(2)    \\
J1850$-$0026 & $--$        & 57708(9)     & 1196.7(2)   & 0.8(2)    \\
J1853+0056 & $--$        & 54970(1)     & 5.4(1)      & 0.4(1)    \\
J1856+0113 & B1853+01 & 57325(34)    & 3489(4)     & 2.8(9)    \\
J1856+0245 & $--$        & 56388(16)    & 2672.5(7)   & 8.4(3)    \\[4.5pt]
J1901+0156 & B1859+01 & 58149(40)    & 1053.02(6)  & 2.1(4)    \\
J1902+0615 & B1900+06 & 55436(6)     & 0.19(3)     & $--$         \\
J1907+0631 & $--$        & 56987(8)     & 2106(2)     & 5.0(5)    \\
J1907+0631 & $--$        & 57901(54)    & 4986(5)     & 4.7(7)    \\
J1909+0749 & $--$        & 57491(9)     & 2904.2(9)   & 3.7(2)    \\[4.5pt]
J1909+0912 & $--$        & 54977(4)     & 4.4(3)      & 0.09(8)   \\
J1909+0912 & $--$        & 57552(60)    & 7389.8(3)   & 4.58(9)   \\
J1913+0446 & $--$        & 55367(1)     & 2.5(2)      & $--$         \\
J1913+0446 & $--$        & 56553(4)     & 0.8(2)      & 0.2(1)    \\
J1913+0904 & $--$        & 56132.8(2)   & 3.2(2)      & $--$         \\[4.5pt]
J1913+1000 & $--$        & 56600(10)    & 491(1)      & $--$        \\
J1921+0812 & $--$        & 55367(33)    & 3632.9(1)   & 3.0(7)    \\
J1952+3252 & B1951+32 & 55328(3)     & 1489.9(4)   & 5.4(4)    \\
J1957+2831 & $--$        & 56897.3(8)   & 2.25(6)     & 1.3(5)    \\
J2005$-$0020 & $--$        & 55976(2)     & 12.5(6)     & $--$         \\[4.5pt]
J2021+3651 & $--$        & 55110(2)     & 2228.8(9)   & 8.5(7)    \\
J2021+3651 & $--$        & 57200(8)     & 3069(2)     & 10.2(9)   \\
J2029+3744 & B2027+37 & 56993.6(5)   & 6.2(2)      & $--$        \\
J2032+4127 & $--$        & 55820(9)     & 273.3(1)    & 1.06(4)   \\
J2219+4754 & B2217+47 & 55857.4(4)   & 1.16(2)     & 0.4(3)    \\[4.5pt]
J2229+6114 & $--$        & 55134(2)     & 198.4(6)    & 6.3(4)    \\
J2229+6114 & $--$        & 55601(4)     & 1223.6(6)   & 13.0(5)   \\
J2229+6114 & $--$        & 56362(5)     & 66.0(9)     & 3.2(5)    \\
J2229+6114 & $--$        & 58424(6)     & 1047.1(9)   & 12.4(6)   \\
J2325+6316 & B2323+63 & 53957(31)    & 0.21(2)     & $-$0.32(4)  \\[4.5pt]
J2346$-$0609 & $--$        & 57495(2)     & 0.55(1)     & 2.4(4)   \\ \hline
\hline
\end{tabular}
\end{scriptsize}
\end{center}
\end{table}
We have identified and measured 106 new glitches in 70 pulsars as part of the Jodrell Bank pulsar timing programme. For this analysis we only consider glitches discovered by JBO which occurred up to 2018 Dec 31, as the sample after this time may be incomplete.  We add these 106 events to the list of published glitches in the Jodrell Bank Glitch Catalogue, forming a total sample size of 543 glitches in 178 pulsars. The glitch epochs and fractional amplitudes in spin and spin-frequency derivative of the 106 new glitches are listed in Table \ref{glitch_table}. Glitch epochs are derived according to the procedure outlined in Appendix \ref{Apn1}.
\begin{figure*}
    \centering
    
    \includegraphics[scale=0.30]{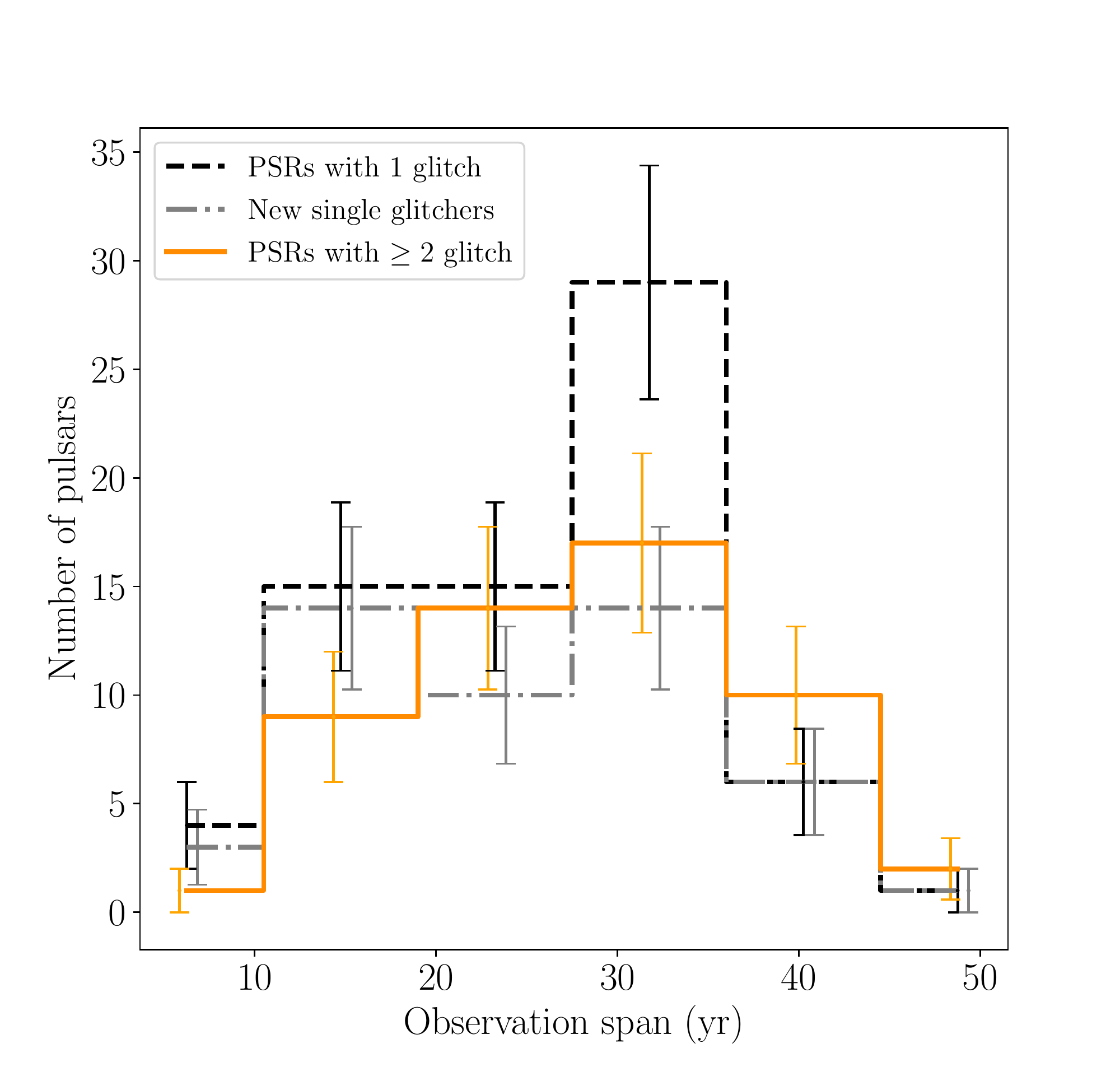}
    \hspace{-0.5cm}
    \includegraphics[scale=0.30]{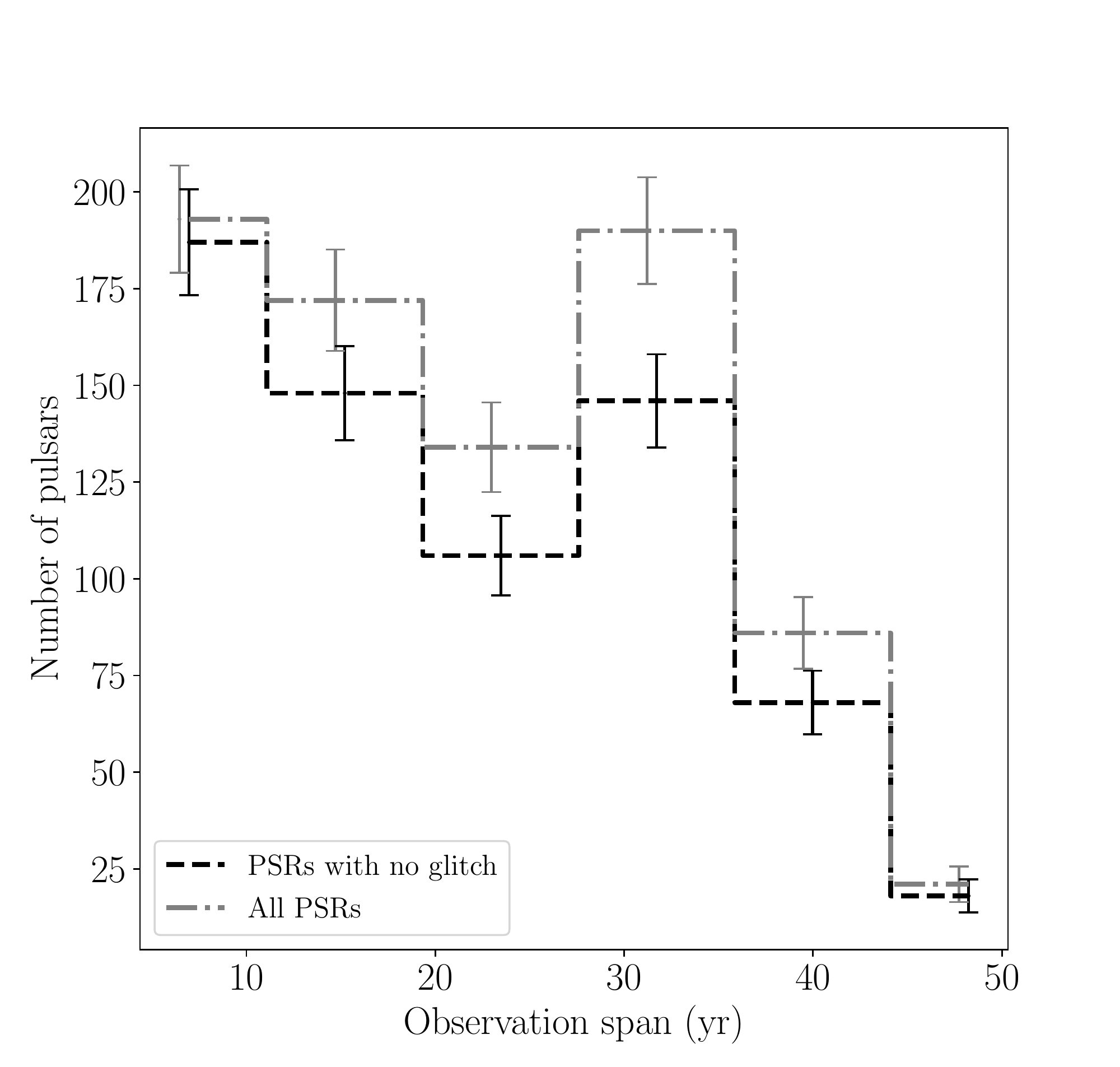}
     \hspace{-0.5cm}
    \includegraphics[scale=0.30]{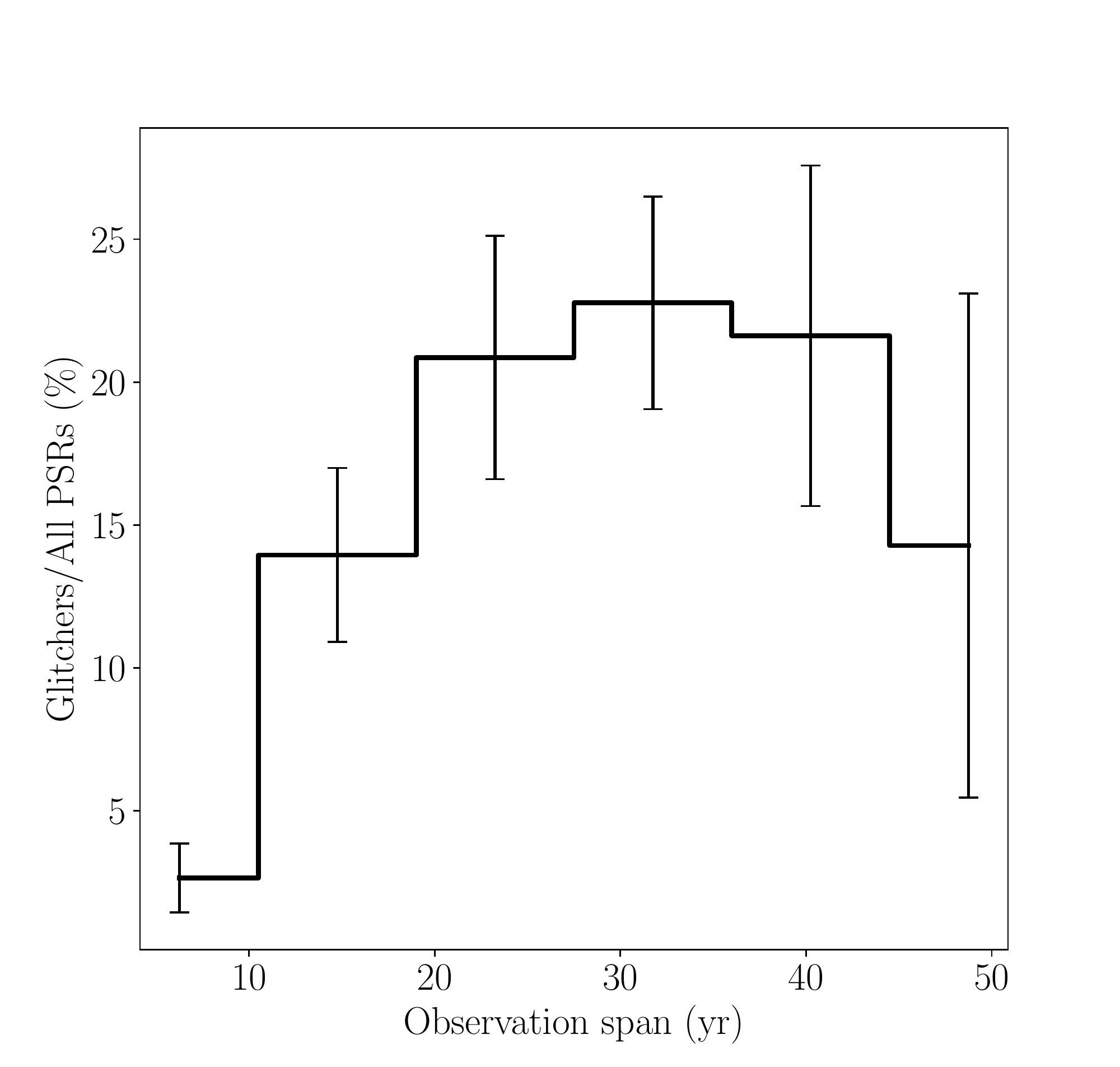}
    
    \caption{The left panel shows the distribution of pulsars belonging to three different categories as function of the time-span covered by their observations. The dashed  and solid lines indicate the pulsars for which one, or more than one glitches are detected respectively. The dash-dotted line shows the distribution of pulsars with only one glitch from our new analysis presented in Table \ref{table_G}. The middle panel shows the corresponding distribution for all pulsars monitored at JBO, and those for which no glitches have been detected till date. The right panel shows the distribution of the percentage of pulsars in which glitches have been detected, as a function of observing span.}
 \label{fig:obsspan}
\end{figure*}

Out of the 70 pulsars for which we report at least one new glitch, a large fraction (70\%) have not previously been observed to glitch. This is much higher than the fraction of glitching pulsars with a single glitch reported in \citet{espinoza2011} (53\%).  Finding more pulsars with a single glitch can be the consequence of the recent addition of pulsars being monitored, as well as of the increased time-span over which pulsars are observed. This is quantified in Figure \ref{fig:obsspan}, which shows the time-span of pulsars observed at JBO for sets of glitching and non-glitching pulsars up to the December 2018 cut-off used here. 

Firstly, the right panel of Figure \ref{fig:obsspan} shows the fraction of pulsars with at least one known glitch as function of the time-span. This demonstrates that pulsars  observed for $\sim10$ yr have a low probability for glitches to be detected. This is a consequence of the typical glitch rate of pulsars, which will be discussed in \S \ref{glitch_rate}. The fraction of pulsars with a single detected glitch increases to $\sim 20$ per cent for pulsars observed for $\geq 20\, \mathrm{yr}$. This fraction should be considered as a lower limit on the intrinsic fraction of glitching pulsars, given that glitches are likely to be discovered in the future for pulsars without any currently known events. Given the glitch rate depends on the spin-parameters (see \S \ref{glitch_rate}), the distribution of monitored pulsars in the $P-\dot{P}$ diagram affects the percentage of pulsars with detected glitches. 

The left panel of Figure \ref{fig:obsspan} shows the distribution of time-spans for the 49 out of 70 pulsars for which we report their first glitch (histogram shown by the dashed-dotted line). Some of these pulsars were not yet observed over a long enough timespan, if observed at all, at the time of the analysis of \citet{espinoza2011}. For others, it is the increased timing baseline which has allowed the discovery of glitches due to their very low glitch rates. This  rate can be as low as one glitch in 48 years of data in the case of PSR J2219$+$4754. In the same panel, the distribution of time-spans for all pulsars with multiple glitches in the JBO data  is shown (solid line histogram). As expected, this distribution is skewed to longer data-sets as these make it more likely for multiple glitches to be detected. 

The shape of the distribution in the left panel depends on the distribution of observing spans for all pulsars monitored at JBO, which is shown in the middle panel ( histogram with dash-dotted line). Many of the pulsars without glitches (histogram with dashed line) have been observed over relatively short spans, hence many more pulsars with reported glitches can be expected in the coming years. There are also pulsars which do not show glitches, despite being observed for decades, suggesting that over time pulsars with even lower glitch rates will be discovered. The glitch-rate and its relation to spin parameters will be further explored in \S ~\ref{glitch_rate}

\subsection{Glitch size ($\Delta \nu$)}\label{jumps}
\begin{figure}
    \centering
    \includegraphics[scale=0.5]{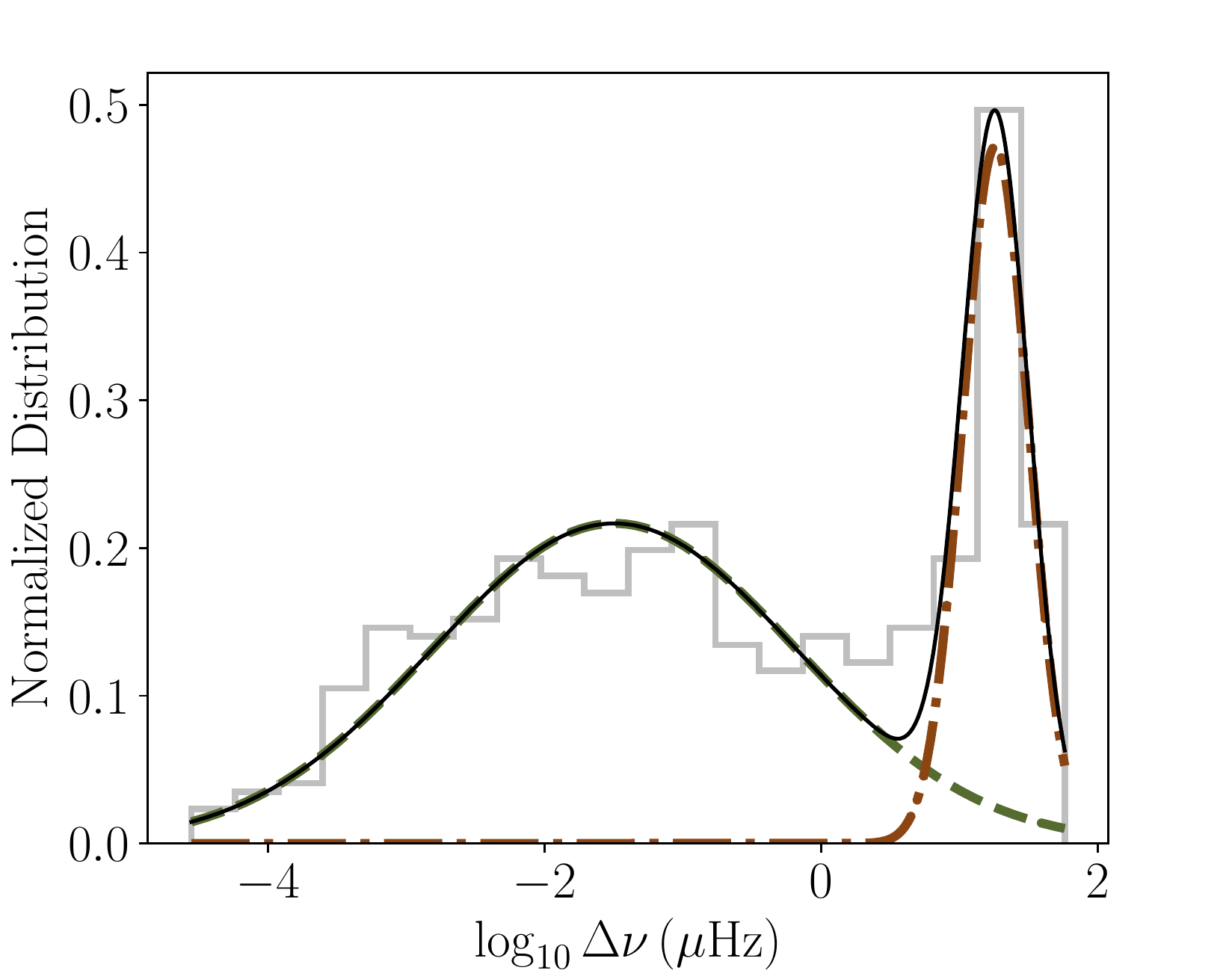}
    \caption{The $\Delta \nu$ distribution for all known glitches (faint grey solid line) overlaid with the results of a Gaussian Mixture Model (GMM). The best model (shown by the solid line) has two Gaussian components (shown by the dashed and dashed-dot line). The narrower component centred at 18 $\mu$Hz represents the large glitches, whereas the wider component centred at 32 nHz represents the smaller glitches.}
    \label{fig:GMM_frequency_jump}
\end{figure}

The magnitude of $\Delta \nu$ indicates the impact of the glitch on the NS crust. In the simplest picture, a larger value indicates a larger amount of angular momentum being transported to the crust. The distribution of $\Delta\nu$ from all known glitches exhibits bi-modality  as reported by \citealt{ Konar+2014} (using 451 glitches in 158 pulsars) and  \citealt{Fuentes+2017} (using 384 glitches in 141 pulsars), using Gaussian Mixture Modelling. We apply the same method to model the distribution of the glitches in our sample as a sum of multiple Gaussian components and find that the distribution can be best modelled by two Gaussian components as shown in Figure \ref{fig:GMM_frequency_jump}. The glitches with small $\Delta \nu$ are modelled by a wide Gaussian component centred at $0.032 \, \mu$Hz with a width of $0.663 \mu$Hz (2$\sigma$ of the Gaussian), whereas the large  $\Delta \nu$ glitches are modelled by a narrow Gaussian component centred at $18 \mu$Hz with a width of $21 \mu$Hz. Potential physical models which explain the origin of bi-modality are discussed by \cite{Celora+2020}.

\begin{figure}
    \centering
    \includegraphics[scale=0.47]{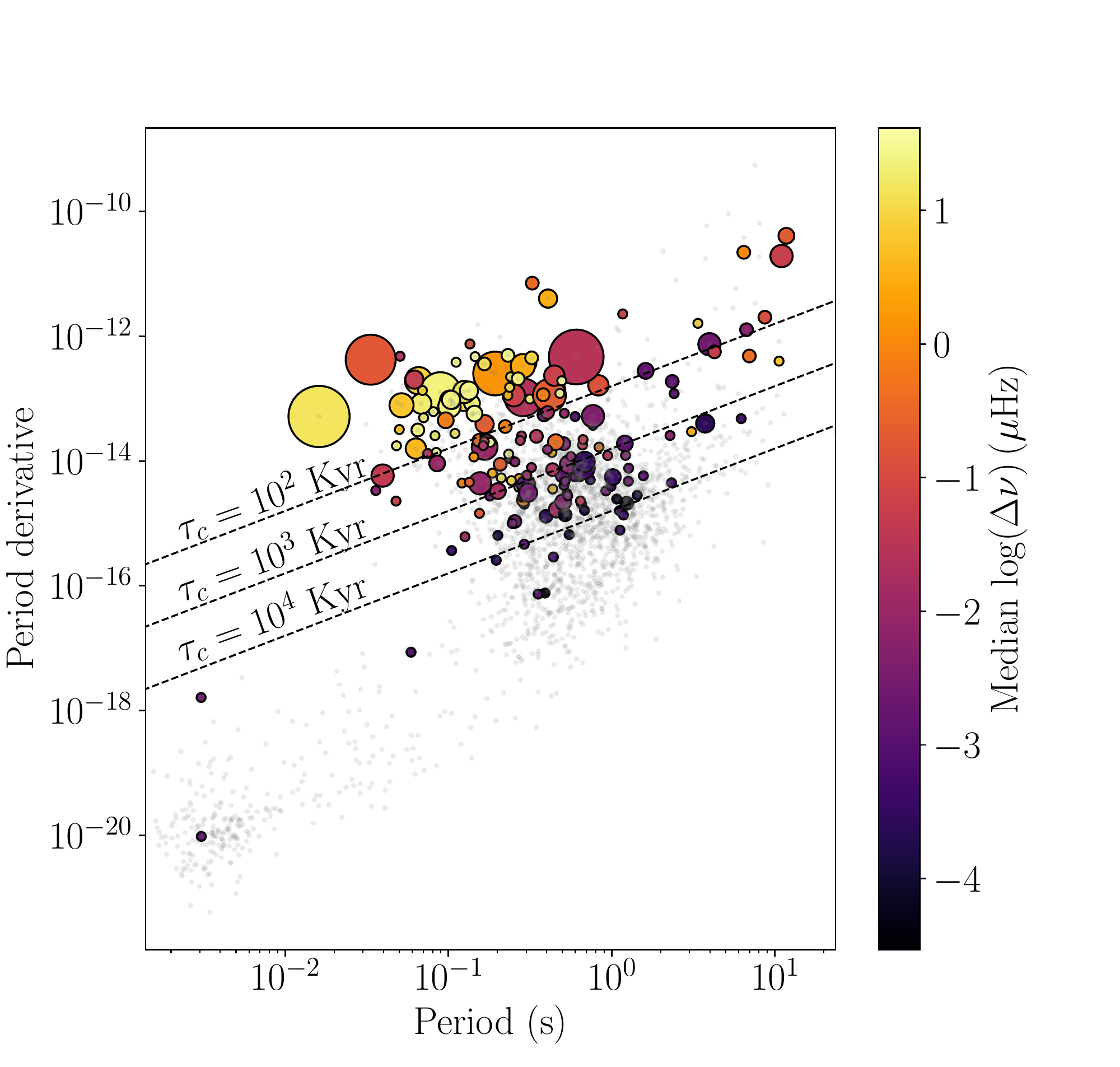}
    \caption{$P - \dot{P}$ diagram showing the known pulsar population (grey points) overlaid with coloured points indicating pulsars with known glitches. The size of the points is proportional to the number of observed glitches. The colour scale shows the median of $\log(\Delta \nu)$ and dashed black lines represent lines of constant characteristic age.}
    \label{fig:P-P_dot}
\end{figure}

To understand the evolution of glitch amplitudes across the pulsar population, we present the variation of the median glitch size and the number of glitches detected in individual pulsars in a $P- \dot P$ diagram in Figure \ref{fig:P-P_dot}.
The colour-scale in Figure \ref{fig:P-P_dot} represents the median glitch size. The size of the points have been scaled to the number of known glitches and it can be observed that more glitches have been discovered in younger pulsars. The median size is less representative of the pulsar's typical $\Delta\nu$ for the rest of the sources, which have few or just one glitch.

Figure \ref{fig:P-P_dot} shows that glitches are predominantly a phenomenon associated with the population of normal pulsars. However two small glitches have been seen so far from  millisecond pulsars:
A micro-glitch with $\Delta \nu/\nu =8(1)\times 10^{-12} $ was detected by \cite{Cognard+Baker+2004} in the millisecond pulsar B1821$-$24, and an even smaller glitch of size $\Delta \nu/\nu =2.5(1)\times 10^{-12}$ in the millisecond pulsar J0613$-$0200  by \citet{Mckee+2016}.
Additionally, several glitches have been observed in magnetically-powered neutron stars (magnetars), sometimes coinciding with emission changes and outbursts \cite[e.g][]{dib08}. Magnetar glitches have typically large $\Delta \nu/\nu$ sizes, owing to their long periods, but exhibit low to intermediate $\Delta \nu$ magnitudes.

\begin{figure*}
    \centering
    \includegraphics[scale=0.75]{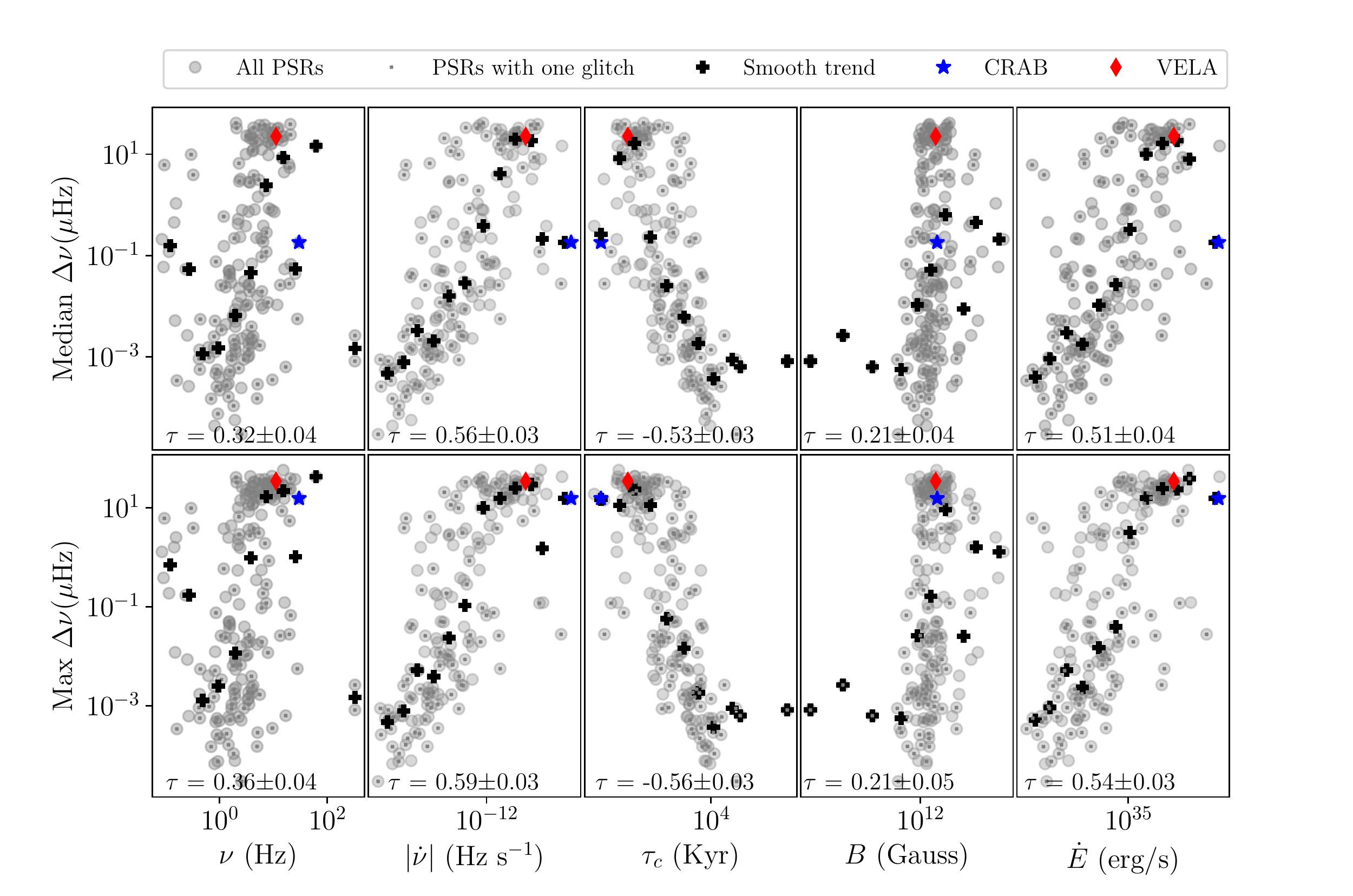}
    \caption{Scatter plots of the median $\Delta \nu$ (upper panel plots) and maximum value of $\Delta  \nu$ (lower panel plots) versus spin frequency $\nu$, spin-down rate $|\dot \nu|$, characteristic age $\tau_c$, inferred dipole magnetic field $B$ and rotational energy loss rate $\dot{E}$. The grey points with a dot in their centre indicate pulsars where only one glitch is detected so far. Black plus markers indicate a smoothed trend to the data obtained by grouping  pulsars in 13 bins based on the independent variables plotted along the horizontal axis. The diamond and star mark represent the Vela and Crab pulsars respectively. We also quote the Kendall-Tau correlation coefficient in every plot. The detailed results for the correlation coefficient are presented in  Table \ref{jump_corr_tab}.}
    \label{fig:glitch_size_correlation}
\end{figure*}

Notably, Figure \ref{fig:P-P_dot} reveals a tendency for larger glitches in younger pulsars, though the apparent trend should be taken with caution as the plotted median does not necessarily represent the typical glitch size -- especially in older pulsars, for which only one or very few glitches are observed (as seen from the point size in the same figure). As the observed tendency is an important result with theoretical implications we looked at the median of the $\Delta\nu$ distribution for each pulsar, and in addition, its maximum observed glitch size.
These parameters are plotted as a function of frequency $\nu$, spin-down rate $\dot{\nu}$, characteristic age $\tau_c$,
magnetic field strength $B$ and the spin-down luminosity $\dot E$ in Figure \ref{fig:glitch_size_correlation}.
We also show the same parameters (maximum and median glitch size) when pulsars are grouped according to $\tau_c$, $\nu$, $\dot{\nu}$, $B$ or $\dot E$ as described below. We observe that the glitch size appears to correlate with the $|\dot \nu|$, as well as with $\tau_c$  and the $\dot E$, whilst no clear trend stands out between the glitch size and $B$ and $\nu$.

In order to quantify any correlations between these parameters, we use the Kendall-Tau test which is suited for data with large scatter, as well as the Spearman's rank correlation test for comparison. The results for the respective coefficients and their probabilities are shown in Table \ref{jump_corr_tab}. 
\begin{table}
\small\addtolength{\tabcolsep}{-3pt}
\begin{tabular}{|c|r|r|r|r|r|}
\hline \hline
Variables & \multicolumn{1}{c}{$\tau$} & $\tau$ (p-value) & \multicolumn{1}{c}{$r_s$} & $r_s$ (p-value) \\ \hline \hline
$\tau$ -vs- Median $\Delta \nu$       & $-0.53\pm0.03$  & $6\times10^{-26}$              & $-0.74\pm0.04$         & $4\times10^{-32}$          \\ 
$|\dot \nu| $ -vs-Median $\Delta \nu$ & $0.56\pm0.03$   & $3\times10^{-28}$              & $0.75\pm0.03$          & $3\times10^{-34}$          \\
$\nu$ -vs-Median $\Delta \nu$         & $0.32\pm0.04$   & $2\times10^{-10}$              & $0.47\pm0.06$          & $2\times10^{-11}$          \\ 
B -vs-Median $\Delta \nu$             & $0.21\pm0.04$   & $5\times10^{-05}$               & $0.30\pm0.06$          & $3\times10^{-05}$           \\ 
$\dot E$ -vs-Median $\Delta \nu$      & $0.51\pm0.04$   & $1\times10^{-23}$              & $0.70\pm0.04$          & $3\times10^{-27}$          \\ 
$\tau$ -vs- Max $\Delta \nu$          & $-0.56\pm0.03$  & $1\times10^{-28}$              & $-0.77\pm0.03$         & $5\times10^{-36}$          \\ 
$|\dot \nu| $ -vs- Max $\Delta \nu$   & $0.59\pm0.03$   & $1\times10^{-31}$              & $0.79\pm0.03$          & $1\times10^{-39}$          \\ 
$\nu$ -vs-Max $\Delta \nu$            & $0.36\pm0.04$   & $9\times10^{-12}$              & $0.50\pm0.06$          & $1\times10^{-12}$          \\
B-vs-Max $\Delta \nu$                 & $0.21\pm0.05$   & $6\times10^{-05}$              & $0.30\pm0.06$          & $2\times10^{-05}$           \\ 
$\dot E$ -vs-Max $\Delta \nu$         & $0.54\pm0.03$   & $6\times10^{-27}$              & $0.74\pm0.04$          & $8\times10^{-32}$          \\ \hline
\hline
\end{tabular}
\caption{The results of the Kendall-Tau ($\tau$) and Spearman's correlation ($r_s$) tests. The first column indicate the variables on which the correlation tests were performed, the second column shows the $\tau$ values with the uncertainties obtained. The third column shows the p-values obtained from the $\tau$ test. The two last columns present the $r_s$ values and its associated p-values.}\label{jump_corr_tab}
\end{table}
In order to obtain a conservative estimate of the correlation coefficient we construct a distribution of the correlation coefficients by performing the Kendall-Tau and the Spearman's rank correlation test 10,000 times (for every pair of variables, e.g median $\Delta \nu$ vs $\tau_c$). For every iteration we sample with replacement an equal number of data points as those present in the actual sample (standard bootstrap re-sampling method) to obtain the standard deviation of the distribution given in Table \ref{jump_corr_tab} as an uncertainty to the correlation coefficients. Whereas, the correlation coefficients were computed from the actual sample.

With the purpose of alleviating any bias that might arise from the small glitch sample of some individual pulsars (generally those with low $|\dot{\nu}|$, but also sources that have not been monitored as long as others), we have grouped pulsars by dividing the $\nu$, $\dot \nu$, $\tau_c$, $B$ and $\dot E$ in 13 equal bins in logarithmic space. We compute the median of each group of pulsars, for the median $\Delta \nu$ (upper panels of Figure \ref{fig:glitch_size_correlation}) and for the maximum $\Delta \nu_{\rm max}$ (lower panels of Figure \ref{fig:glitch_size_correlation}). This results in the crosses in Figure \ref{fig:glitch_size_correlation}. We further mark the position of the Crab and Vela pulsars in all these plots as they are the most studied pulsars in terms of glitching behaviour.

\begin{figure}
    \centering
    \includegraphics[scale=0.44]{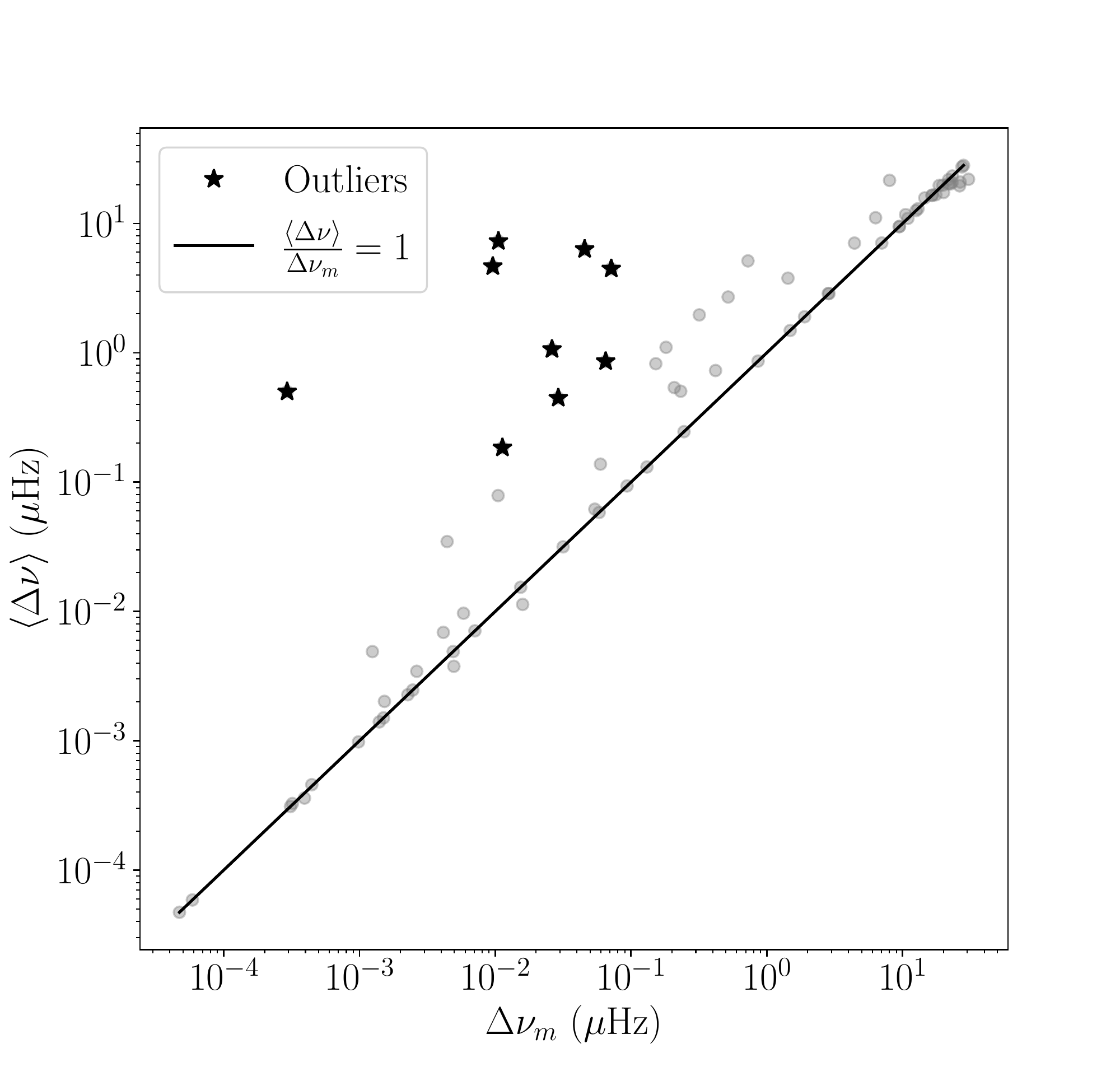}
    \caption{The average glitch size $\left<\Delta \nu\right>$ versus the median glitch size $\Delta \nu_m$ of 80 pulsars, shown as grey points. The black line follows the relation $\left<\Delta \nu\right> = \Delta \nu_m$. The star marked points indicates the 9 pulsars with a skewed $\Delta \nu$ distribution (see the text for detailed discussion).}
    \label{fig:median_and_average_size_plot}
\end{figure}

From the smooth trend we can see some hint for a turnover taking place: the median glitch size increases with increasing $|\dot{\nu}|$ (correspondingly, decreasing $\tau_c$ since $\tau_c \propto \nu/ |\dot\nu|$), reaching a peak for pulsars with parameters similar to the Vela pulsar. For higher $|\dot{\nu}|$ (correspondingly, lower $\tau_c$), i.e. Crab-like pulsars, the median size decreases again. Such an turnover is not  prominent when looking at the maximum $\Delta \nu_{\rm  max}$-vs-$|\dot{\nu}|$ or $\Delta \nu_{\rm max}$-vs-$\tau_c$ curves. We believe this feature stems from the fact that pulsars whose glitch behaviour is similar to the Crab, show a wider range of glitch sizes, often skewed to smaller sizes, which results in lower median sizes (\citealt{Melatos+2008,JRF+2019}, see also Figure \ref{fig:median_and_average_size_plot}). On the other hand, pulsars like the Vela have more of a ``typical" glitch size (so the average and median are closer together). 

Besides the correlations with $|\dot \nu|$ and $\tau_c$, we further find generally positive correlations between glitch size and the $\dot E$, except at the highest values of $\dot E$  where a turnover occurs (reflecting the turnover in $\Delta\nu$-vs-$|\dot \nu|$) and the correlation becomes negative. We also find a generally poor correlation between glitch size and $\nu$. Similarly, there is little overall correlation between glitch size and $B$ (though with the exception of Vela, there may be some evidence of correlation in the 'normal' pulsars) , however as Figure \ref{fig:glitch_size_correlation} shows, the few pulsars with the weakest inferred $B$ (which include the two millisecond pulsars) have so far only been seen to exhibit small glitches.

As seen in Table \ref{jump_corr_tab}, the correlation of both the median and maximum $\Delta\nu$ appears the strongest with $|\dot \nu|$. The results for the other parameters are not then surprising, considering they are inferred from the rotational parameters and their scaling with the spin-down rate is $\tau\propto|\dot{\nu}|^{-1}$, $\dot{E}\propto\dot{\nu}$ and  $B\propto|\dot{\nu}|^{1/2}$. However, some of these parameters can also be expected to independently show some correlation with glitch size, as for example the spin-up amplitude  might depend on the neutron star's internal temperature (see e.g. \cite{HA14}), which in turn will be a function of both its $\tau_c$ and $B$. This can also explain the hint seen in Figure \ref{fig:glitch_size_correlation} towards a decrease in glitch size (both median and maximum) for neutron stars with the highest $B$ (e.g. magnetars), as these will be in general hotter than their same age counterparts in the typical pulsar population.

To investigate the $\Delta \nu$ distribution for individual pulsars we construct Figure \ref{fig:median_and_average_size_plot} which shows the relation between the median $\Delta \nu \, (\Delta \nu_m)$ and average $ \Delta \nu \, (\left<\Delta \nu\right>)$ for a given pulsar, which reveals the skeweness of its glitch size distribution. Pulsars with symmetrical glitch size distributions (e.g. Gaussian) should fall on the straight line which corresponds to $\left<\Delta \nu\right>/ \Delta \nu_m=1$. We consider outliers to be those that lie more than 1 standard deviation from the relation and they are marked with stars in Figure \ref{fig:median_and_average_size_plot}. For all of the outlier pulsars $\left<\Delta \nu\right> > \Delta \nu_m$, indicating a tail in the distribution towards large $\Delta \nu$ (like, for example, in the PSR J1740-3015 which in general shows small/intermediate size glitches but also occasional large events).

We note that many old (low $|\dot{\nu}|$) pulsars have just one detected glitch, and so the median and maximum $\Delta\nu$ are the same (hence excluded from our analysis presented in Figure \ref{fig:median_and_average_size_plot}). Therefore we cannot be certain how these results would change if we observed over enough time to see multiple glitches from these sources. There are, however, enough Vela-like pulsars with multiple glitches, to suggest the possibility of an evolutionary trend in glitch size, size distribution and overall glitch activity.

\subsection{Spin-down rate changes ($\Delta \dot \nu$)}\label{delta_nu_dot_sec}
\begin{figure}
    \centering
     \includegraphics[scale=0.43]{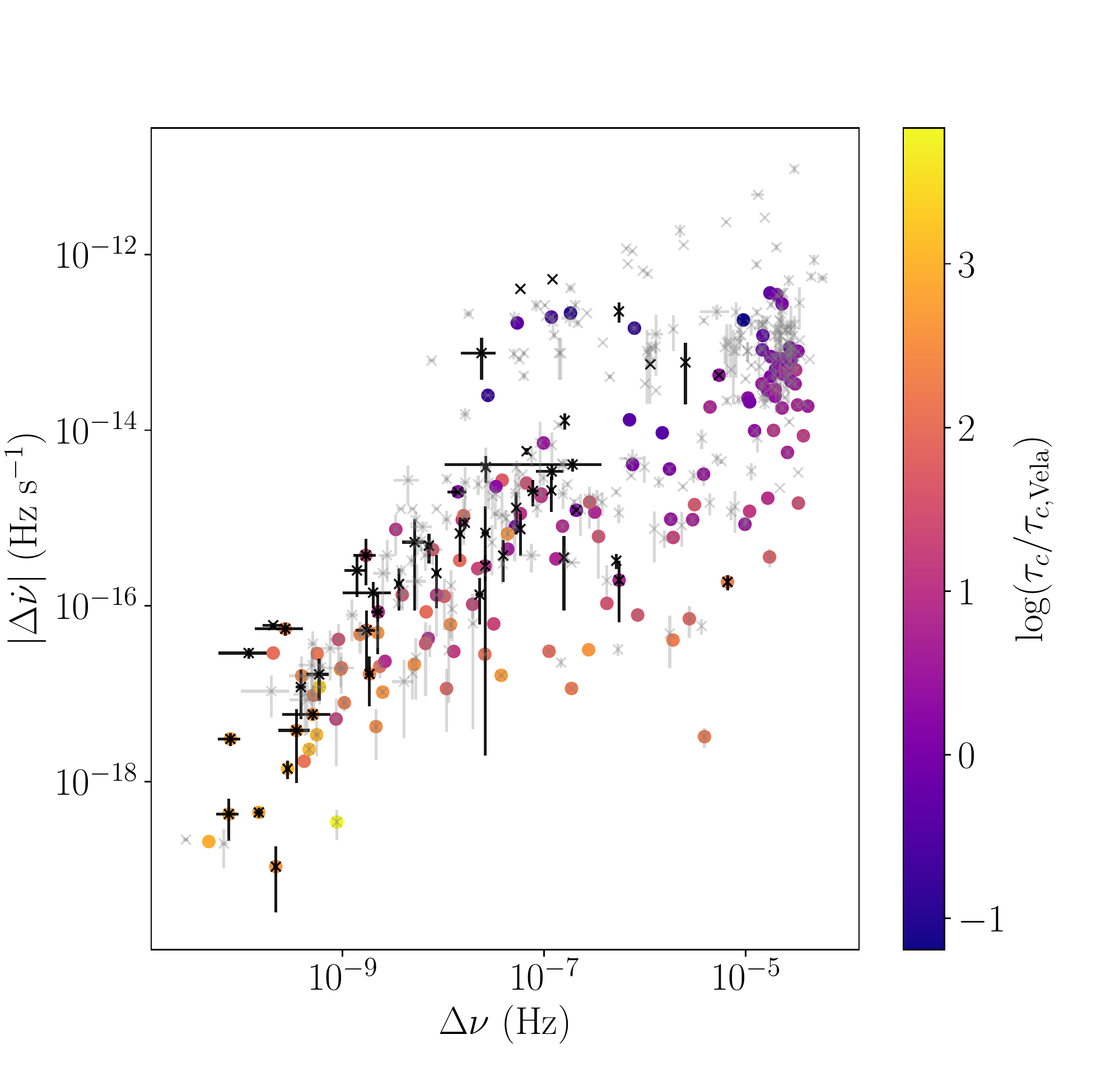}
    \includegraphics[scale=0.47]{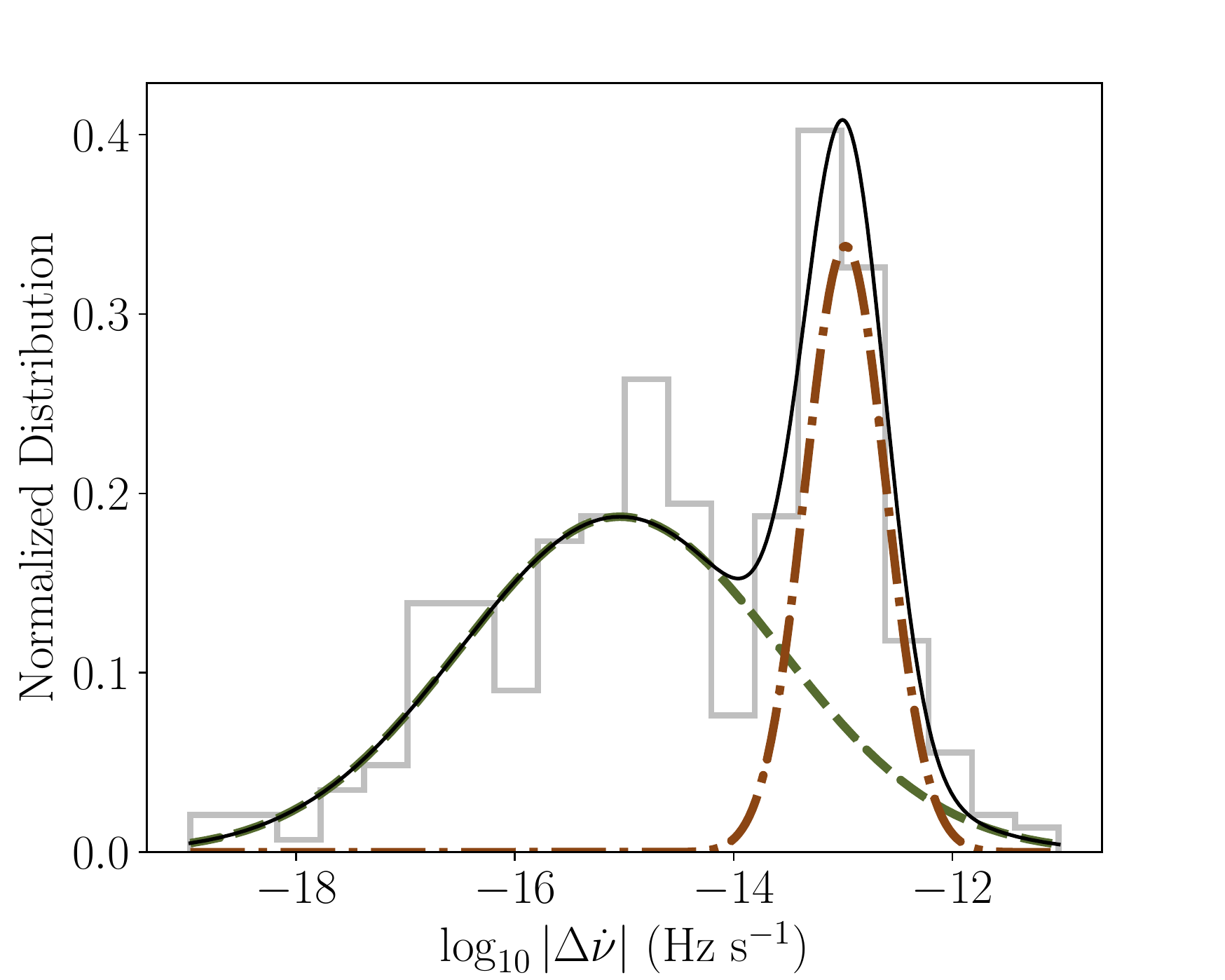}
    \caption{The upper panel  shows glitches in the parameter space $\Delta \nu$--$|\Delta \dot \nu|$. The black points indicate $\Delta \dot \nu > 0$ and all grey crosses indicate ``canonical" glitches with $\Delta \dot \nu < 0$. The coloured points show the median values of $\Delta \nu$ and $\Delta \dot \nu $ for every individual pulsar with the colour-scale indicating the ratio of the $\tau_c$ of the pulsar to the $\tau_c$ of the Vela pulsar. The lower panel shows the Gaussian Mixture Model (GMM) for the $|\Delta \dot \nu|$ distribution. The  component denoted by the dashed-dot line models glitches where large changes in the spin-down rate were measured, centred at $1.0 \times 10^{-13}$~Hz~s$^{-1}$ and $2.0\times10^{-13}$~Hz~s$^{-1}$ wide. Component denoted by the dashed line includes glitches with smaller values of $\Delta \dot \nu$ and is centred at $9 \times10^{-16}$ Hz s$^{-1}$ and is $266\times10^{-16}$ Hz s$^{-1}$ wide.}
    \label{fig:delta_nu_dot_plots}
\end{figure}

At a glitch, the change in the rotation frequency is often accompanied by a change in the spin-down rate $\Delta \dot \nu$, as can be seen in Figure \ref{fig:delta_nu_dot_plots}. The inferred change can be either positive or negative, however in the majority of glitches negative values of $\Delta \dot \nu$ are seen, indicating an increased spin-down rate following the spin up. In order to investigate the nature of the distribution of $\Delta \dot \nu$, we perform a Gaussian Mixture modelling on the measured values of $\Delta \dot \nu$. We find that this distribution, as was the case for the glitch size distribution (Figure \ref{fig:GMM_frequency_jump}), is best modelled by the sum of two Gaussian components, shown in Figure \ref{fig:delta_nu_dot_plots}. The smaller values of $\Delta \dot \nu$ are modelled by the wider Gaussian component and the larger values by the narrower component. Such a similarity in the nature of the two distributions is due to the observed correlation\footnote{Note that there are several observational biases at play which can give rise or amplify this correlation, we therefore do not attempt to quantify it.} between $\Delta \nu$ and $\Delta \dot \nu$. This can be seen in the upper panel of Figure \ref{fig:delta_nu_dot_plots}, where we have plotted the absolute value of $\Delta \dot \nu$ versus $\Delta \nu$ for all measured glitches except where the change in spin-down rate is consistent with zero. The plot clearly indicates that larger glitches are associated with larger changes in the spin-down rate. We also observe a clustering of points at the right hand top corner of the same figure, which contributes to the narrower Gaussian component in both the distribution of $\Delta \nu$ and $\Delta \dot \nu$. In order to investigate the clustering we plot the median $\Delta \nu$ and median $\Delta \dot \nu$ for every individual pulsar (including the sources with only one glitch) on the upper panel of Figure \ref{fig:delta_nu_dot_plots}. The colour-scale indicates the ratio of the $\tau_c$ of the pulsar to the $\tau_c$ of Vela in log-scale. We note that the clustered, large, glitches come from pulsars with ages very close to the age of the Vela pulsar, possibly indicating an evolutionary link as already discussed when considering glitch sizes.  

\subsection{Glitch Rate}\label{glitch_rate}

\begin{figure}
    \centering
    \includegraphics[width=\columnwidth]{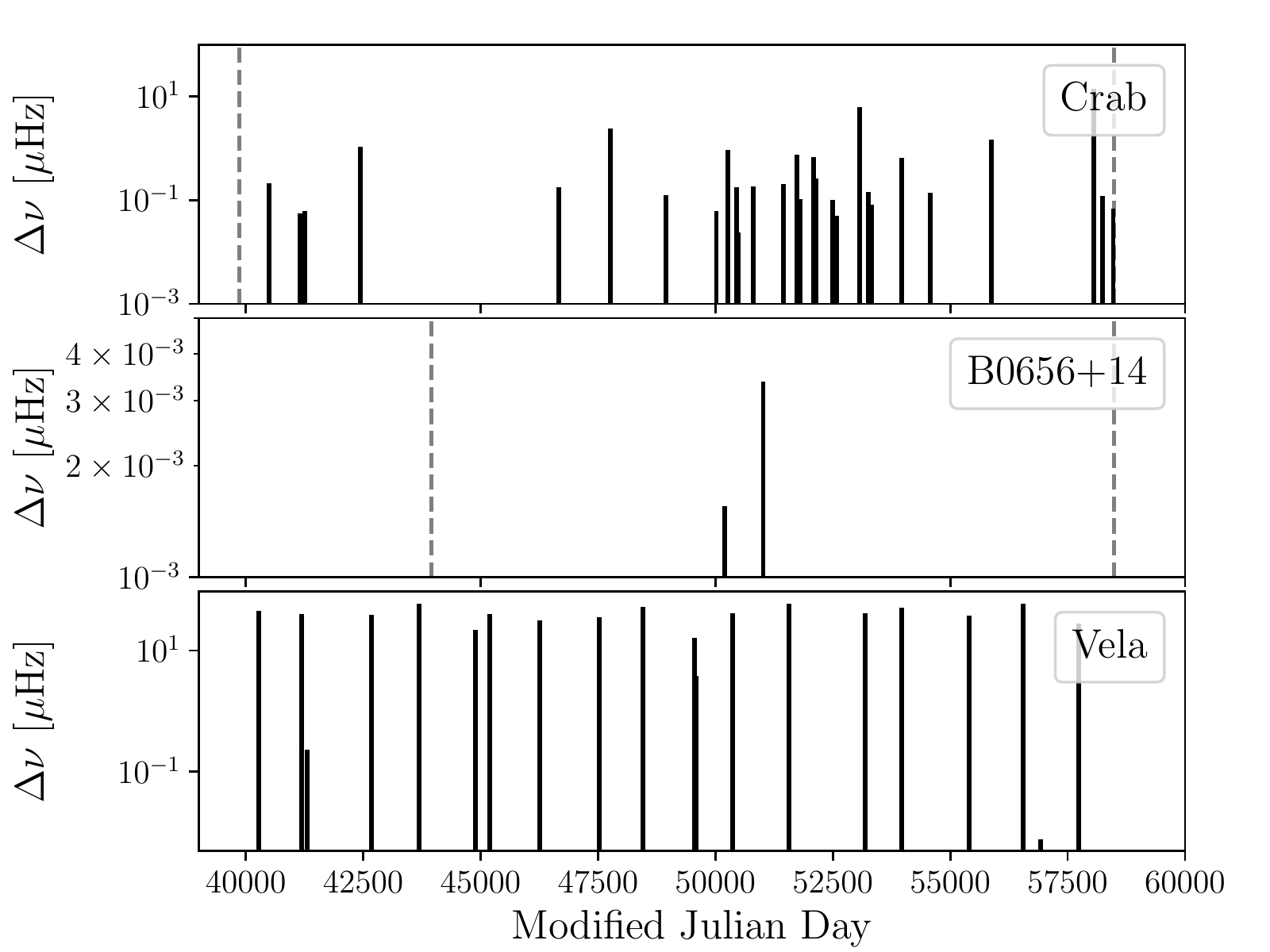}

    \caption{The sequence of glitches and their sizes $\Delta \nu$ as a function of time for the Crab pulsar, PSR B0656$+$14, and the Vela pulsar. The glitch occurrence times for the Crab pulsar and PSR B0656$+$14 are irregular, but frequent in the former and rare in the latter. The glitch sizes for these pulsars also vary considerably. Conversely, the Vela pulsar exhibits glitches of similar sizes with a more consistent inter-glitch interval. The vertical dashed lines indicate the time span over which these pulsars have been monitored at JBO.}
    \label{fig:example_irregular_glitch}
\end{figure}

Different pulsars exhibit different kinds of glitch behaviour, as already discussed with regard to their glitch size distributions, but this is also true in the rate at which they glitch. For example the time intervals between glitches for the Crab pulsar are  irregular, whereas in other pulsars, such as  Vela, glitches appear more equally spaced in time. Whilst the glitch rate is a useful metric to get an idea of how active a pulsar is in terms of glitches, one should keep in mind that this rate might not be constant over the years of observations. In fact, it has been shown that the Crab pulsar had statistically significant increased activity between 1995 and 2006 \citep{Lyne+2015, Carlin+2018}, as apparent in Figure \ref{fig:example_irregular_glitch}. 

    The glitch rates we present here, calculated under the assumption that they are constant in time, should therefore be viewed only as approximations. They are also sensitive to the timespan over which they have been calculated: in this study we consider the average glitch rate for the full interval over which a pulsar has been monitored for glitches. If instead one were to take the interval between the first and last detected glitches, it would overestimate the glitch rate. This is illustrated for PSR B0656+14 in Figure \ref{fig:example_irregular_glitch}; it is clear that without information on the time period before and after glitches --over which the pulsar was actually monitored but no glitches were discovered-- the inferred glitch rate would have been greatly overestimated. We therefore confine the sample of pulsars used for glitch rate calculations, to those observed at the JBO, for which we have adequate information on the time intervals over which these pulsars have been monitored and do not risk introducing this bias. This sample spans the full $P-\dot{P}$ diagram (see Figure \ref{fig:Jodrell_psr}), and with 134 pulsars a large, representative, fraction of the known glitching pulsars is included in the analysis.

\begin{figure}
    \centering
    \includegraphics[scale =0.45]{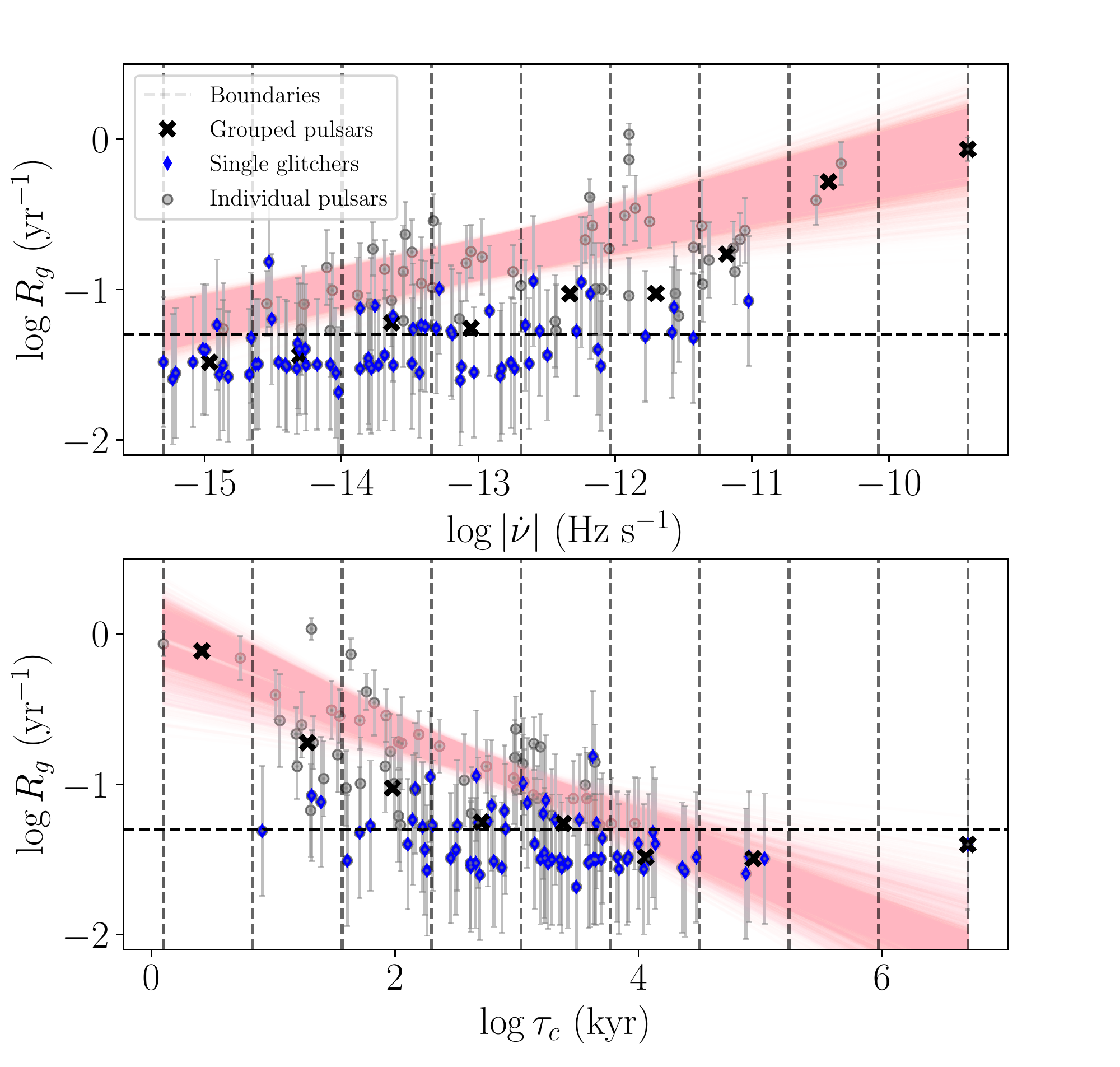}
    \caption{Dependencies of the glitch rate on the magnitude of the spin-down rate $|\dot{\nu}|$ (upper panel) and characteristic age $\tau_c$ (lower panel). The grey circles correspond to individual pulsars and the black crosses indicate the median glitch rate for the group of pulsars bounded by the dashed grey vertical lines. The diamonds indicate the pulsars with only one known glitch.  The shaded region show the distribution of probable power law models describing the data with a glitch rate greater than 0.05 yr$^{-1}$, (horizontal dashed line).}
    \label{glitch_rate_dep}
\end{figure}

The glitch rate is defined as 
\begin{equation}\label{rate}
    R_{g} = \frac{N}{T},
\end{equation}
where $N$ is the total number of observed glitches in time interval $T$ of observations at the JBO, computed by taking the difference between the first and the last observed ToA  in our dataset for every individual pulsar. By approximating glitches to be a Poisson process the uncertainty on the glitch rates were computed as $\sqrt{N}/T$.
The minimum glitch rate found was 0.02 yr$^{-1}$ and the maximum 1.07 yr$^{-1}$. The change in glitch rate as a function of $|\dot{\nu}|$ and $\tau_c$ can be seen in Figure \ref{glitch_rate_dep}. We present values for individual pulsars (grey circles), as well as the median glitch rate (black crosses) for groups of pulsars within the intervals of $\tau_c$ and $|\Dot{\nu}|$ shown by the vertical grey lines. We verified that the use of the group average glitch rate, instead of the group median, has no qualitative effect on our results.

In Figure \ref{glitch_rate_dep} pulsars with a single detected glitch are indicated using the diamond symbols. We note that the single-glitch pulsars are concentrated towards the low glitch-rate end of the distribution, however they do not appear to be exceptional in terms of their  $\tau_c$ and $|\Dot{\nu}|$.

 We also attempt to model the glitch rate 
as a function of $\tau_c$ using a power law (PL) model of the form $A\tau_c^\alpha$, as has been explored in the literature before (see e.g. \citet{espinoza2011}). We consider only pulsars with  $R_g > 0.05$ yr$^{-1}$ as the extrapolation to lower rates is greatly hindered by the total observing time. In order to arrive at  estimates of $A$ and $\alpha$ and their uncertainties, we utilise a standard bootstrapping method (as described earlier). The process was iterated over $10^{4}$ times to obtain the distribution of the model parameters. The best fit values are $A= 1.0^{+0.2}_{-0.2}$ and $\alpha= -0.29^{+0.03}_{-0.02}$, the upper and lower bounds in the parameters corresponds to the 68\% confidence interval. The reduced chi-square ($\chi^2_\mathrm{red, PL}$) is 1.7(5)\footnote{The number inside the parenthesis indicates the 1$\sigma$ error}. To assess the significance of the $\tau_c$ dependence, the PL model was compared to a model which is constant, and equal to the weighted mean (WM) of the glitch rate for the same samples used for PL model fits.
The reduced chi-square ($\chi^2_\mathrm{red, WM}$) in this case is 4(1), which indicates that the PL model is favoured over the WM model, which implies that the glitch rate decreases with $\tau_c$. The distribution of possible power law models is shown by the shaded region in Figure \ref{glitch_rate_dep}. 

We also explore the possible dependencies between the glitch rate and the spin-down rate by fitting a power law of the form $B|\dot \nu|^\beta$. The best-fit values of the parameters are $B=76^{+37}_{-39}$ and $\beta=0.187^{+0.020}_{-0.007}$.  The distribution of models is shown by the shaded region in the upper panel of Figure \ref{glitch_rate_dep}.  As above, we also tried a model with a constant value computed from the weighted mean. We find $\chi^2_\mathrm{red, PL} =2.1(8)$ and $\chi^2_\mathrm{red, WM} =4(1)$, therefore, as before, the power-law dependence is marginally favoured over one where the glitch rate is independent of $|\dot \nu|$. 
An increasing glitch rate with the magnitude of the spin-down rate is in accordance with the notion that glitches are driven by the spin-down. Assuming all glitches relate to the pinning and unpinning of superfluid vortices, the trend can be understood in terms of the underlying mechanism invoked for generating stress between the superfluid and non-superfluid components, i.e. an increasing lag between the two. In other words, pulsar with lower $\dot{\nu}$ needs more time to develop the velocity lag required to trigger a glitch. The same argument holds if the trigger of unpinning is a crustquake because the stress builds in the lattice due to spin-down (e.g., \cite{haskellmelatos15}).  

Our observed trend of lower glitch rate with increasing $\tau_c$ is consistent with the results obtained by \citet{McKenna+Lyne+1990, espinoza2011}. 
The internal temperature likely plays a role \citep{McKenna+Lyne+1990}, as the higher temperatures of younger pulsars facilitate creep and unpinning of vortices \citep{AlparvortexcreepI1984}. Therefore, glitches might be triggered at a smaller velocity lag between the superfluid and normal component when the pulsar is hotter. In this case it might also be expected that their glitches will be smaller, potentially explaining the lower value of median $\Delta \nu$ in the very young, Crab-like, pulsars (as shown in Figure \ref{fig:glitch_size_correlation}).

\begin{table}
\begin{center}
\begin{tabular}{|l|c|c|c|c|}
\hline \hline
Variables           & $\tau$ & $\tau$ (p-value) & $r_s$ & $r_s$ (p-value) \\ \hline \hline
$\dot \nu$-vs-$R_g$ & $-0.37$  & $5.1\times10^{-10}$                                                   & $-0.55$ & $4.1\times10^{-11}$                                                  \\
$\tau_c$-vs-$R_g$   & $-0.41$  & $1.2\times10^{-11}$                                                   & $-0.58$ & $9.4\times10^{-13}$                                                  \\ \hline \hline
\end{tabular}
\caption{Correlation analysis between the glitch rate $R_g$ and both $\dot{\nu}$ and $\tau_{c}$. The first column indicates the variables between which the correlation coefficients were calculated. The second and third column indicates the Kendall-tau correlation coefficient and the its p-value respectively. The fourth and fifth column show Spearman's correlation coefficient and its p-value respectively.}\label{corr_glt_rate}
\end{center}
\end{table}

\begin{figure}
    \centering
    \includegraphics[scale=0.45]{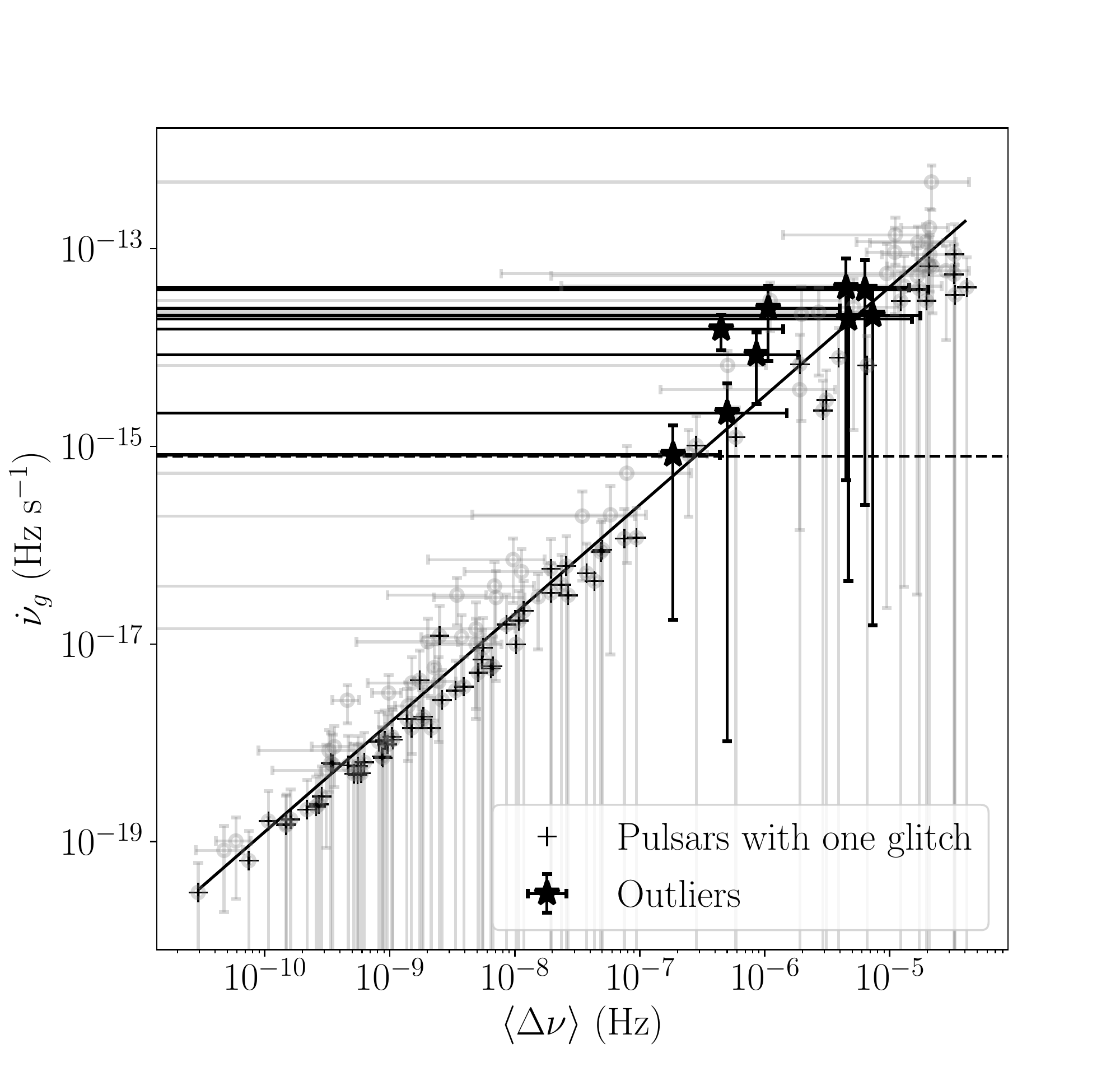}
    \caption{Variation of the glitch activity as a function of average glitch size for all known glitching pulsars (grey points). The star marked points represent the 9 pulsars which have a skewed $\Delta \nu$ distribution (see the text for detailed discussion). The plus marked points are the 70 pulsars with only one detected glitch. The black line is a linear fit to all points within the parameter space. The gradient of the line is 1.1(1). The black horizontal line corresponds to a transition point between low and high glitch activity as obtained from the minimum of the GMM fit shown in Figure \ref{fig:GMM-activity}.}
    \label{fig:nu_g_delta_nu_corr}
\end{figure}

\begin{figure}
    \centering
    \includegraphics[scale=0.48]{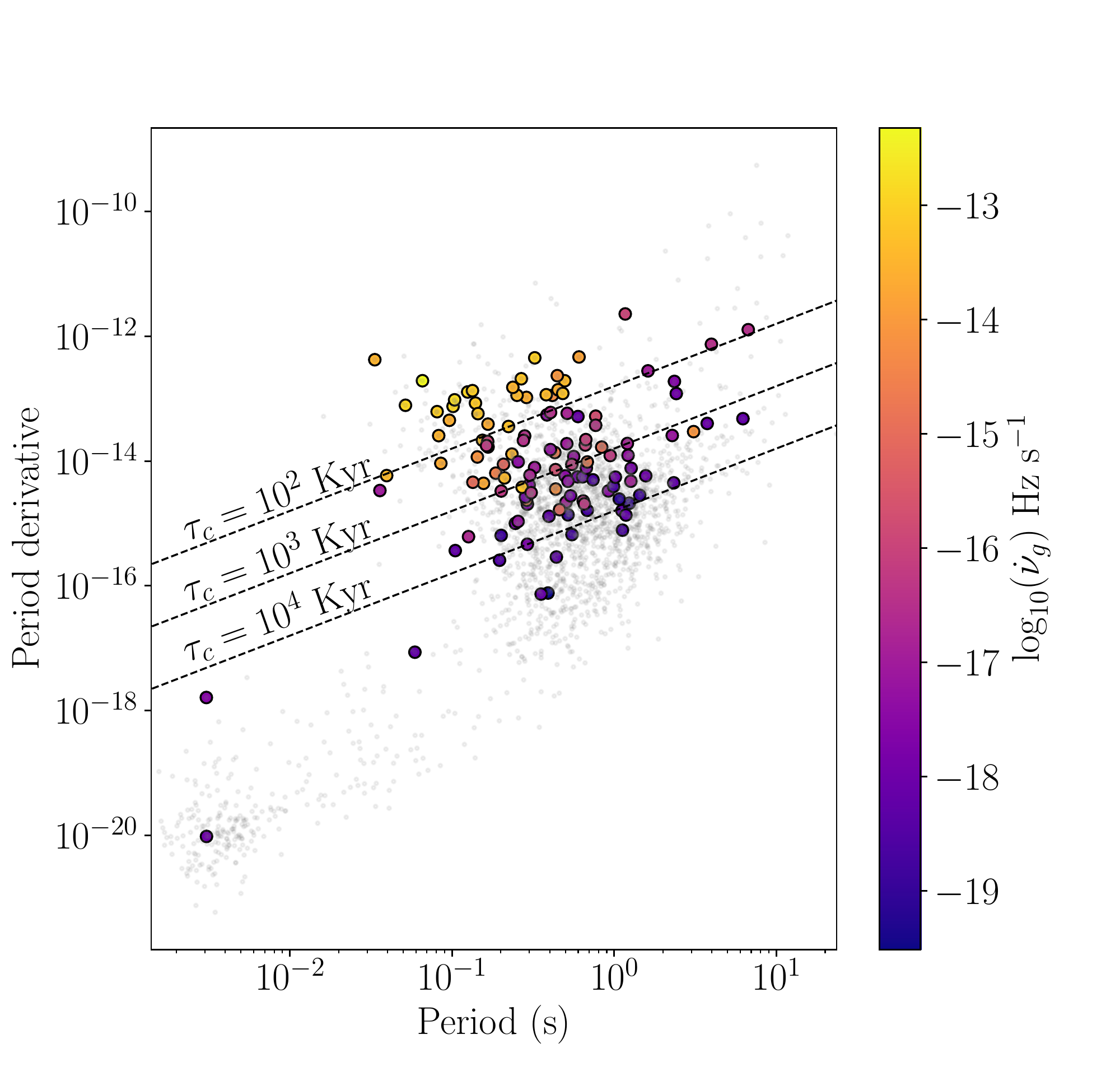}
    \caption{$P-\dot P$ diagram showing all known pulsars (faint grey points) overlaid with glitching pulsars for which glitch activity could be calculated (see text). The colour bar represents the glitch activity on a logarithmic scale.}
    \label{fig:ppdotactivity}
\end{figure}

\begin{figure}
    \centering
    \includegraphics[scale=0.44]{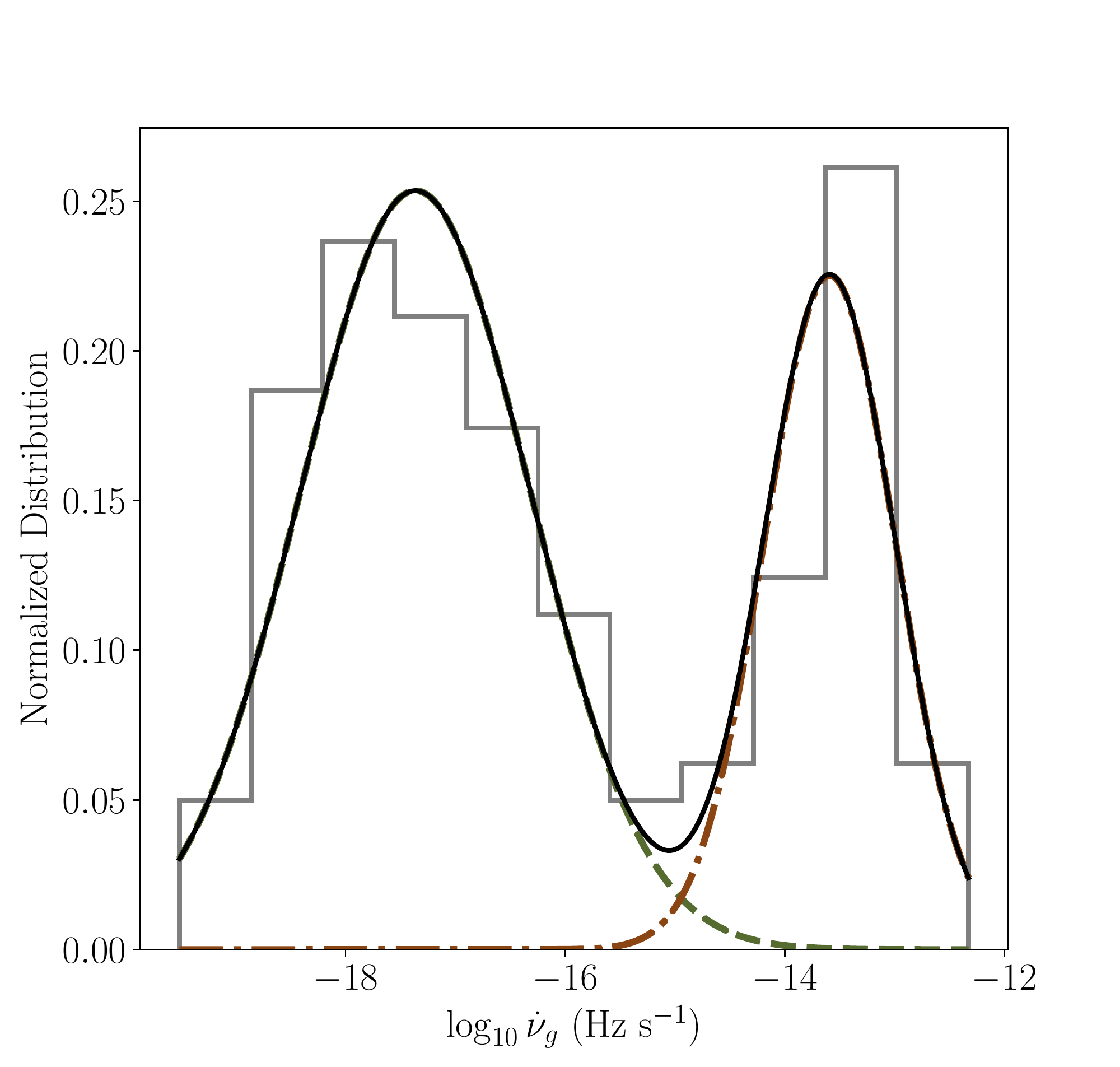}
    \caption{A GMM showing a bimodal distribution of glitch activity. The black solid line indicates the combined GMM model. The dashed and the dashed-dotted lines indicate the individual components.}
    \label{fig:GMM-activity}
\end{figure}

\begin{figure}
    \centering
    \includegraphics[scale=0.46]{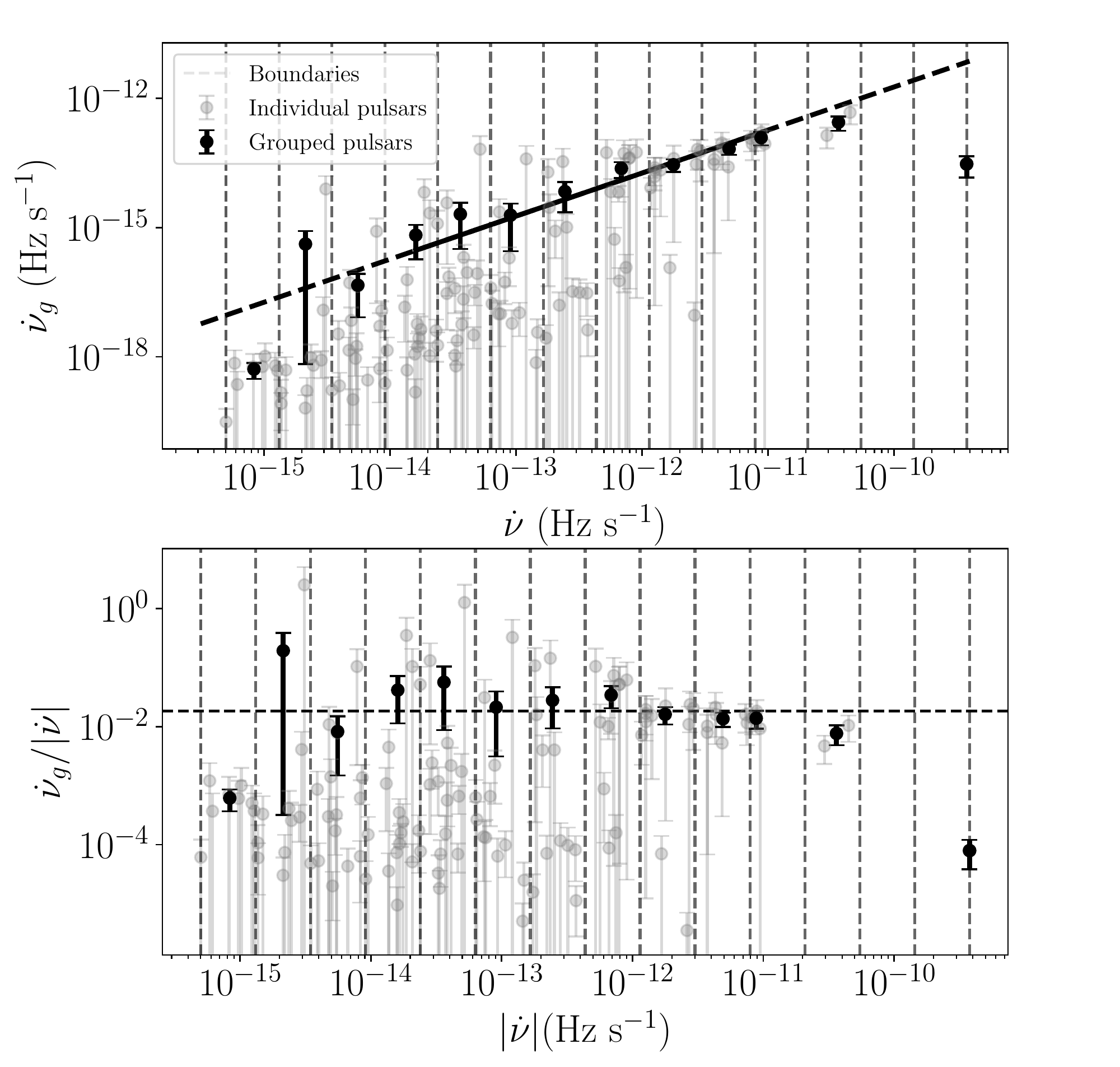}
    \caption{Variation of the glitch activity (upper panel) and  $\dot \nu_g/ |\dot \nu|$ (lower panel) as a function of spin-down rate. The grey points shows the glitch activities of individual pulsars and the black points represent the average glitch activity of pulsars grouped within ranges of $|\dot \nu|$ denoted by the vertical grey dashed lines. The black solid line shows the straight line fit ($\dot \nu_g/ |\dot \nu| = 0.0018(3)$) to the data with $-14 < \log \dot \nu < -10.5$. The horizontal line corresponds to $\dot \nu_g/ |\dot \nu| = 0.0018$.}
    \label{fig:activityplot}
\end{figure}

\subsection{Glitch activity}\label{glitchactivity}
Pulsars emit particle winds and electromagnetic radiation at the cost of their rotational energy, which eventually leads to a secular spin-down. However, due to glitches, a certain fraction of the crustal spin-down gets reversed. A measure of this is the glitch activity ($\dot \nu_g$), defined as the cumulative spin-up due to multiple glitches over a total observation span $T$ \citep{Lyne+shemar+smith+2000, espinoza2011, Fuentes+2017}:
\begin{equation}\label{single_pulsar_glt_act_eq}
    \dot \nu_{g} = \frac{\sum_k \Delta \nu_k}{T}.
\end{equation}
Here $\Delta \nu_k$ is the change in the spin frequency at the $k$-th glitch and $T$ is the duration over which the pulsar has been  monitored. We adopt the same methodology to compute $T$ as discussed in  \S \ref{glitch_rate}. The uncertainty on the glitch activity is given by the relation  $\sigma_{\dot \nu_g} = T^{-1} \sqrt{\sum_k \Delta \nu_k^2}$ \citep{Fuentes+2017}. The glitch activity can be expressed as a product of the glitch rate and the average glitch size since
\begin{equation}\label{glitch_act_withavg_size}
        \dot \nu_g = \Big(\frac{N}{T}\Big)\Big( \frac{\sum_k \Delta \nu_k}{N}\Big)= R_g\left<\Delta \nu\right>.
\end{equation}
Hence, we can expect some correlation between activity and average glitch size, which is confirmed in Figure \ref{fig:nu_g_delta_nu_corr}. It is clear that the ``outliers" (for which median glitch size is smaller than the average, see Figure \ref{fig:median_and_average_size_plot} and related text) generally exhibit  a high glitch activity and have relatively large average glitch sizes ($\left<\Delta \nu\right>\gtrsim 0.1 \;\mu\text{Hz}$).

The dependence of the glitch activity on the spin parameters of a pulsar is shown in the $P-\dot P$ diagram in Figure \ref{fig:ppdotactivity}.  It can be seen that younger pulsars tend to have a higher glitch activity than older pulsars. We show a histogram of glitch activity in Figure \ref{fig:GMM-activity} that shows a clear bi-modal distribution.  In order to quantify this we perform a GMM analysis, finding that a two component Gaussian best models the data, as expected.  The broader peak is centred at $4.0\times10^{-18}$ Hz s$^{-1}$ and is $5.0\times10^{-17}$ Hz s$^{-1}$ wide. Comparing Figure \ref{fig:ppdotactivity} and \ref{fig:GMM-activity} we identify that the broader Gaussian component represent the older population of pulsars. The higher glitch activity of the younger population of pulsars (indicated by the darker points in Figure \ref{fig:ppdotactivity}) are represented by the component centred at $2.0\times10^{-14}$ Hz s$^{-1}$ with a width of $9.0\times10^{-14}$ Hz s$^{-1}$. The minimum of the cusp between the two Gaussian occurs at $8.0\times10^{-16}$ Hz s$^{-1}$ indicating a transition from the low glitch activity to the high glitch activity. It is worth noting that, as the glitch activity is proportional to $\left< \Delta \nu \right>$ and rate $R_g$, and in turn these are proportional to $|\dot{\nu}|$ (on which characteristic age depends), the bi-modalities seen in these parameters are different realisations of the same underlying phenomenon.

\begin{figure}
    \centering
    \includegraphics[scale=0.40]{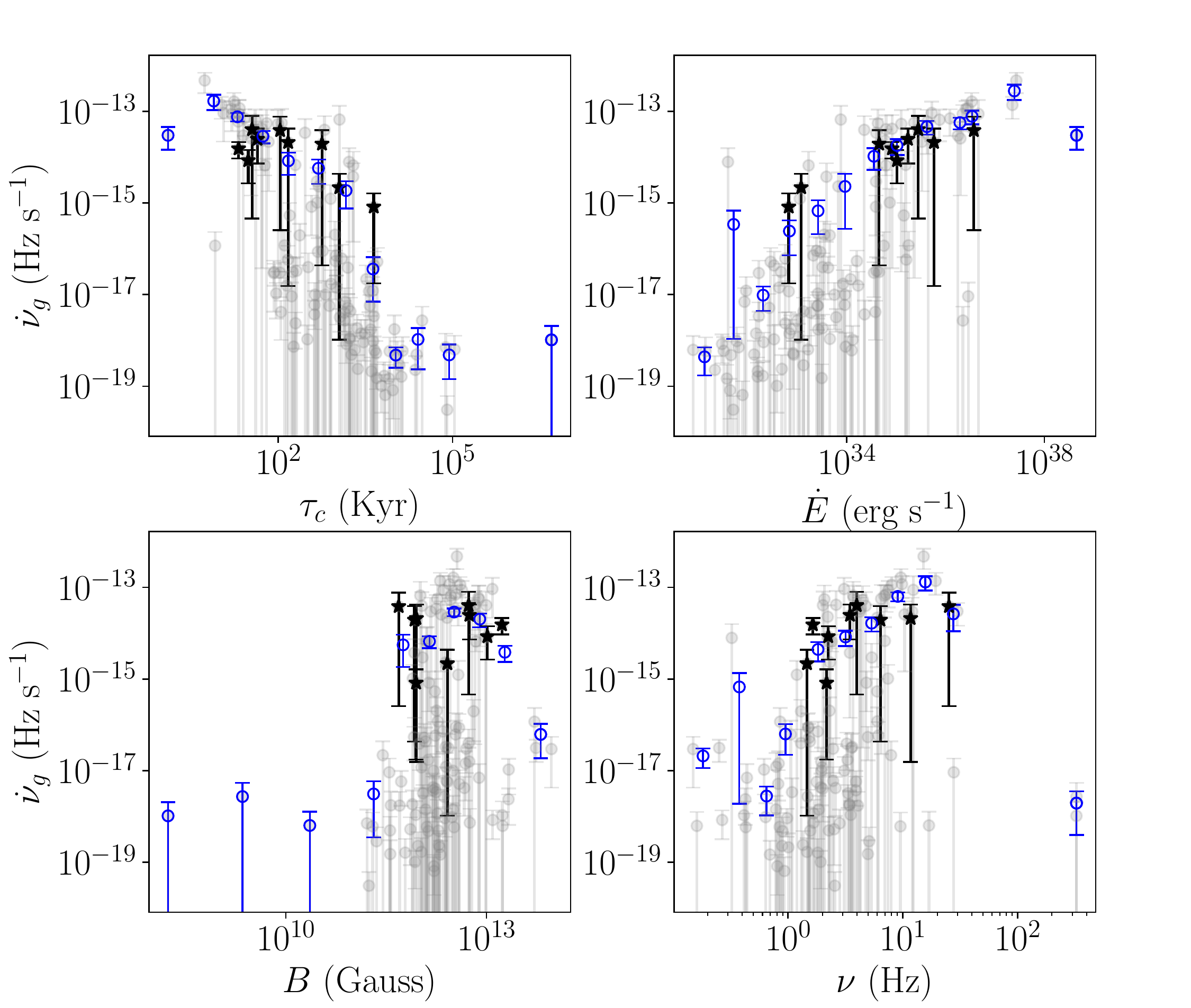}
    \caption{Glitch activity as a function of various measured and derived pulsar parameters. The grey points represent glitching pulsars monitored as part of the JBO pulsar timing programme, whilst the hollow circles are the averaged activity over groups of pulsars binned according to the parameter shown on each horizontal axis. The star marked points are outliers (see text for details).}
    \label{fig:activity_otherparam}
\end{figure}

In the literature, the average glitch activity is often computed by grouping pulsars according to  parameters like $\dot \nu, \tau_c $, etc. Following \citep{Lyne+shemar+smith+2000} the average glitch activity in a group is defined as
\begin{equation}\label{average_glt_act}
    \bar{\dot{\nu}}_g = \frac{\sum_j \sum_k \Delta \nu_{kj}}{\sum_j T_j},
\end{equation}
where the index $j$ is over the pulsars in the group and index $k$ is over all the glitches of a given pulsar in the group. Figure \ref{fig:activityplot} shows the glitch activity as a function of $\dot \nu$. The grey points are the glitch activity for individual pulsars and the black points are the average glitch activity computed using Equation \ref{average_glt_act} by grouping the pulsars according to their $\dot \nu$. The uncertainties on the average glitch activity is given by $\sigma_{\bar{\dot{ \nu}}_g} =T^{-1}\sqrt{\sum_j \sum_k \Delta \nu_{jk}}$ \citep{Fuentes+2017}.  \citet{Lyne+shemar+smith+2000}, \citet{espinoza2011} and \citet{Fuentes+2017} found (each using the glitch sample available at the time) that  $\dot \nu_g \propto |\dot \nu|$ in the range $-14 < \log \dot \nu < -10.5$. The slope of a straight line fit to the $\dot \nu_g $ and $\dot \nu$ (shown by the black solid line in Figure \ref{fig:activityplot}) in the given range is $\dot \nu_g/|\dot \nu| = 0.018\pm0.003$ (or $\log_{10}(\dot \nu_g/|\dot \nu|)=-1.74\pm0.07$), implying on average only 1.8\% of the pulsar's spin-down is reversed by glitches. This can be used to constrain the moment of inertia of the angular momentum reservoir ($I_{\text{res}}$) for the pulsars exhibiting regular glitches. Following \citep{Link+1999}, one can approximate $I_{\text{res}}/I_{\text{c}} = \dot \nu_g / |\dot \nu|$, where $I_{\text{c}}$ is the moment of inertia of the crust and any stellar component tightly coupled to it. The lower panel of Figure \ref{fig:activityplot} shows this fraction of the reversed spin-down as a function of the spin-down rate, and the horizontal line indicates 1.8\% reversed spin-down rate. However, it should be noted that the $1.8\%$ spin-reversal is only valid for the pulsars where we have observed at least one or more glitches over the time span of the monitoring programme. We confirm the results of \citet{espinoza2011} and \citet{Fuentes+2017} that the pulsars at the two extremes, with $|\dot \nu| < 10^{-14}$ Hz s$^{-1}$ and $|\dot \nu| > 10^{-10.5}$ Hz s$^{-1}$, deviate from the horizontal line, showing a smaller glitch activity, though it should be noted that in the sample we consider only the Crab pulsar falls in the highest $|\dot{\nu}|$ bin.  

We further explore the dependence of the glitch activity as a function of other pulsar parameters obtained from the different combinations of $\nu$ and $\dot \nu$. Figure \ref{fig:activity_otherparam}, shows that glitch activity decreases with characteristic age, as expected like we already discussed.  We note that the pulsars whose glitch size distribution is skewed tend to have lower characteristic ages.  Similarly as $\dot E ( \propto \nu \dot \nu)$ and $|\dot \nu|$ are correlated positively with glitch size (and rate), activity tends to be also greater for pulsars with greater energy loss rates. We also note that the pulsars with the lowest magnetic field strength ($B \propto \nu^{1/2}\dot\nu^{-3/2}$) have correspondingly low glitch activity (see also Figure \ref{fig:activity_otherparam}), and that a decrease in glitch activity is evident in the most strongly magnetised neutron stars. However in both these cases, there are relatively few events therefore it is difficult to draw any robust statistical conclusions. 

\section{Conclusions}\label{conclude}
The JBO timing programme routinely monitors 800 pulsars from which we have identified 106 new glitches in 70 pulsars. These include many pulsars which have been observed to glitch only once, since the increased timing baseline allowed the detection of glitches in pulsars with a very low glitch rates. Our analysis was restricted to the data set acquired up to and including December 2018. Using our updated sample of glitches, we have examined the overall statistics of the glitching pulsar population. The findings from our analysis are summarised below:
\begin{itemize}
   \item Glitch sizes ($\Delta \nu$) are generally smaller in pulsars with low energy loss rates (i.e. low $\dot{\nu}$ and inferred $\dot{E}$) and older characteristic age, as confirmed from our correlation analysis. There is no clear correlation with spin-frequency and magnetic field strength, although pulsars with low magnetic fields strengths which have been observed to glitch have only exhibited comparatively smaller glitches, and very  highly-magnetised neutron stars have not so far  displayed very large glitches.
   \\
   \item The overall glitch size distribution is bi-modal in nature, and as $\Delta\nu$ is further correlated with the measured $\Delta \dot{\nu}$ this leads to a  bi-modality in the distribution of $|\Delta \dot \nu|$. Within the bi-modal size distribution we find that the component corresponding to larger glitches is mainly due to Vela-like pulsars which tend to exhibit glitches of a characteristic large size.
   \\
    \item Pulsars with low spin-down rates tend to exhibit longer time intervals between glitches (i.e., they have a lower glitch rate), which is qualitatively consistent with a model in which glitches are driven by the spin-down of the neutron star. Glitches are rarer in  pulsars with large characteristic age. Such a connection has been confirmed by the correlation analysis between the glitch rate and both the spin-down rate and characteristic age.
    \\
    
    \item Due to its dependence on glitch size, glitch activity is also bi-modally distributed, wherein pulsars with higher glitch activity tend to be younger than those with low activity. Despite the relatively broad observed range of glitch rates, we find that a pulsars' glitch activity correlates strongly with their average glitch size. On average, 1.8\% of the spin-down rate is reversed due to glitches in pulsars where $-14 < \log \dot \nu < -10.5$.
\end{itemize}

We continue to monitor more than 800 pulsars as part of the JBO timing programme, and we further continue to search for new glitch events.

\section*{Data Availability}

The data underlying the work in this paper are available upon reasonable request.

\section*{Acknowledgements}
Pulsar research at Jodrell Bank is supported by a consolidated grant from the UK Science and Technology Facilities Council (STFC). D.A. acknowledges support from an EPSRC/STFC fellowship (EP/T017325/1). 

\appendix

\section{Uncertainty on the glitch epoch}\label{Apn1}

\begin{figure}
    \centering
    \includegraphics[scale=0.4]{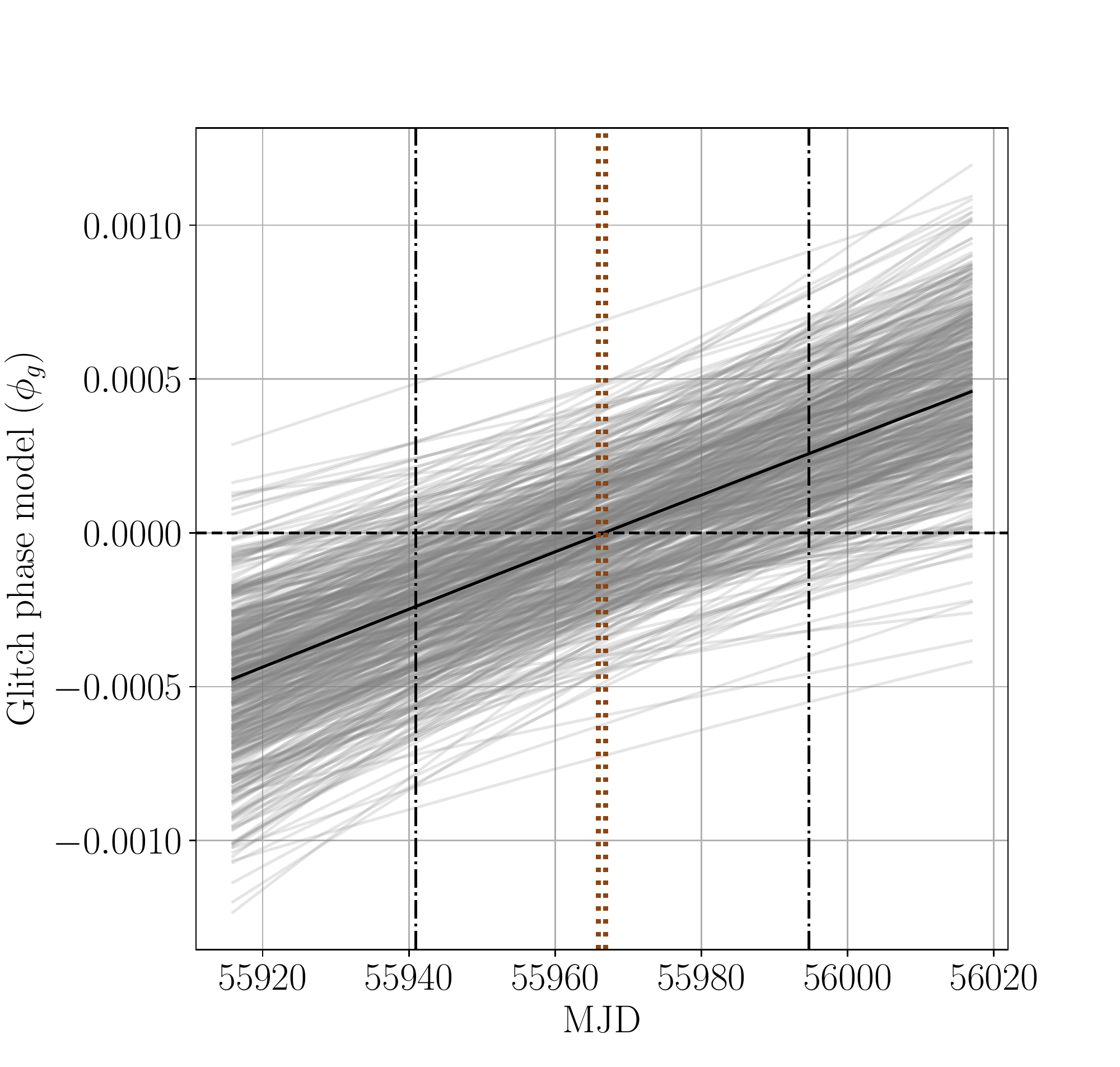}
    \caption{
    This figure corresponds to the glitch model of PSR J0215+6218 on MJD 55966. It shows the time evolution of the glitch phase. The black solid line indicates the average of the models and grey lines indicates its distribution. The dotted lines indicate the ToAs just before and after our initial estimate of $t_g$ and the dashed-dotted lines indicate the range of possible glitch epochs.
    In this case $\phi_g$ is close to zero for all plausible glitch models.
    }
    \label{fig:GELP_0}
\end{figure}

\begin{figure}
    \centering
    \includegraphics[scale=0.4]{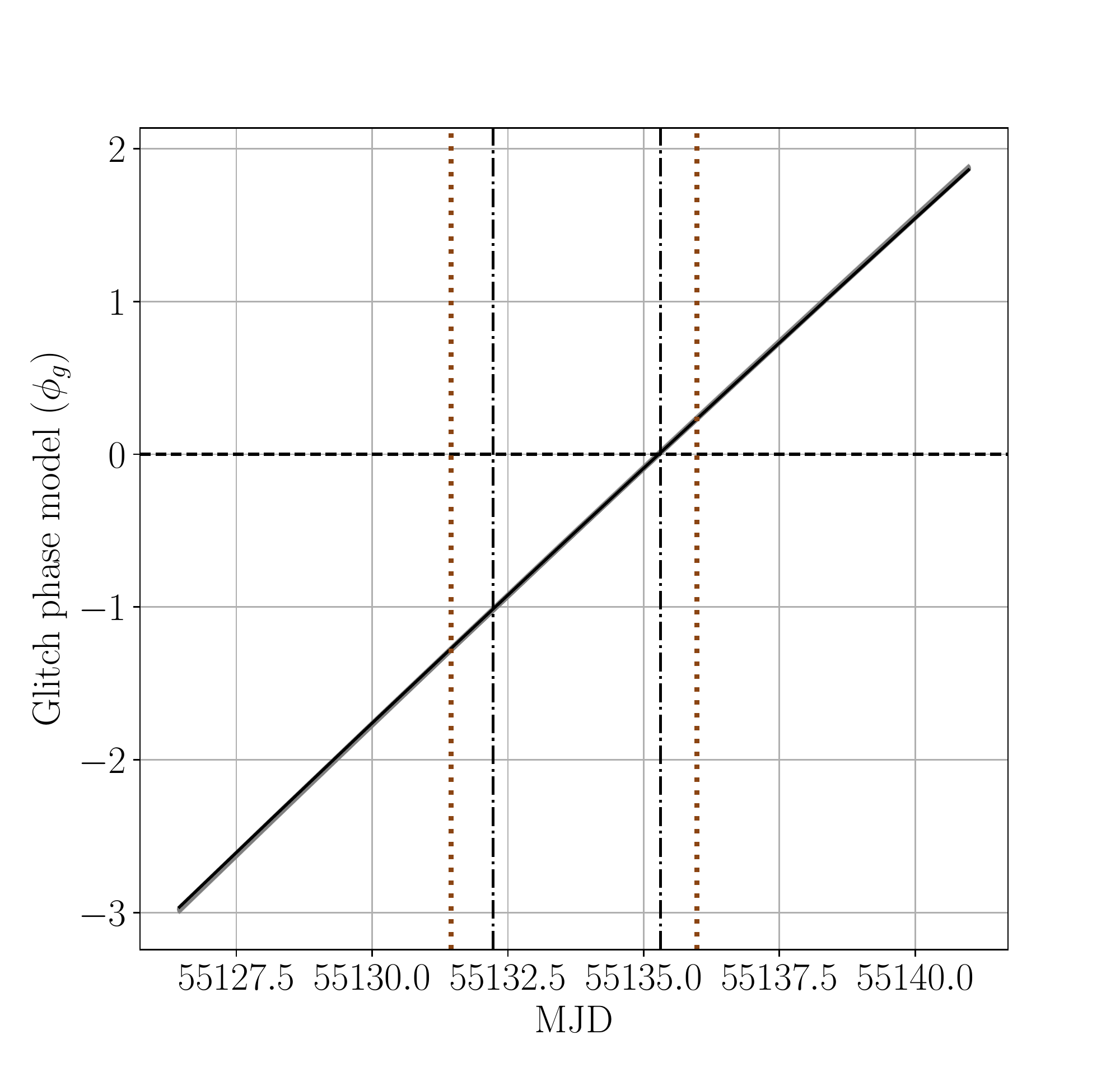}
    \caption{This figure is similar to figure \ref{fig:GELP_0}, but corresponds to the glitch of PSR J2229+6114 on MJD 55134. Here multiple integer values of $\phi_g$ are possible, which defines the uncertainty on the glitch epoch as indicated by the dash-dotted lines. See appendix \ref{Apn1} for details.
    }
    \label{fig:GLEP_many}
\end{figure}

Depending on the glitch amplitude and the observing cadence, an ambiguous number of pulsar rotations can occur between the last pre-glitch observation and the first post-glitch observation. Therefore all solutions for the time $t_g$ which lead to $\phi_g$ (which quantifies the effect of the glitch, see Equation \ref{glitch_evolution}) having an integer value at a time between the two ToAs bracketing the glitch are valid solutions for the glitch epoch. 
For this reason there may be multiple solutions for $t_g$. In order to estimate a glitch epoch and a corresponding uncertainty we use the measured glitch parameters (discussed in the \S \ref{timing}) and their $1\sigma$ uncertainties to construct a Gaussian distribution, from which we randomly sample to obtain a list of roots for $t$ where $\phi_g(t)$ has integer values. For smaller glitches, the initial guess of $t_g$ is more uncertain, so the solution $\phi_g(t)=0$ might be found at $t$ outside the interval defined by the ToAs bounding the initial $t_g$ estimate.
In those cases the mean of the roots is taken to be $t_g$, and the standard deviation is quoted as the uncertainty on the glitch epoch. An example of such a scenario is presented in Figure \ref{fig:GELP_0}.
For larger glitches, the roots bounded by the ToA just before and after our initial estimate of $t_g$ corresponds to different integer values of $\phi_g(t)$. We select the times corresponding to the extremes of roots within the ToA bounds. The midpoint of these times 
is taken to be an estimate of $t_g$ and the uncertainty is taken to be half of the total time span between these extreme values. Figure \ref{fig:GLEP_many} shows an example of this scenario.

\bibliographystyle{mnras}
\bibliography{reference.bib} 
\end{document}